\documentclass[prl,superscriptaddress,reprint,amsmath,amssymb,aps,longbibliography]{revtex4-1}
\usepackage{graphicx}
\usepackage{float}
\usepackage[
pdfstartview=FitH,
bookmarks=true,
breaklinks=true,
colorlinks=true,
linkcolor=blue,anchorcolor=blue,
citecolor=blue,filecolor=blue,
menucolor=blue,urlcolor=blue]{hyperref}

\usepackage{xcolor}
\usepackage{ulem}

\begin{document}

\title{Observation of modulation-induced Feshbach resonance}

\author{Tongkang Wang}
\affiliation{Beijing Academy of Quantum Information Sciences, Beijing 100193, China}
\affiliation{Department of Physics and State Key Laboratory of Low Dimensional Quantum Physics, Tsinghua University, Beijing, 100084, China}
\author{Yuqi Liu}
\affiliation{Department of Physics and State Key Laboratory of Low Dimensional Quantum Physics, Tsinghua University, Beijing, 100084, China}
\author{Jundong Wang}
\affiliation{Department of Physics and State Key Laboratory of Low Dimensional Quantum Physics, Tsinghua University, Beijing, 100084, China}
\author{Youjia Huang}
\affiliation{Department of Physics and State Key Laboratory of Low Dimensional Quantum Physics, Tsinghua University, Beijing, 100084, China}
\author{Wenlan Chen}
\email{cwlaser@ultracold.cn}
\affiliation{Department of Physics and State Key Laboratory of Low Dimensional Quantum Physics, Tsinghua University, Beijing, 100084, China}
\affiliation{Frontier Science Center for Quantum Information and Collaborative Innovation Center of Quantum Matter, Beijing, 100084, China}
\author{Zhendong Zhang}
\email{zhendongzhang19950715@gmail.com}
\affiliation{Department of Physics and Hong Kong Institute of Quantum Science and Technology, The University of Hong Kong, Hong Kong, China}
\author{Jiazhong Hu}
\email{hujiazhong01@ultracold.cn}
\affiliation{Beijing Academy of Quantum Information Sciences, Beijing 100193, China}

\begin{abstract}
In this work, we observe a novel resonant mechanism, namely the modulation-induced Feshbach resonance. By applying a far-detuned laser to the cesium $D_2$ transition with intensity modulation, we periodically modulate the energy levels of atomic collisional states.
This periodic modulation connects the free-scattering states to shallow molecular states. At specific frequencies, significant atom loss is observed, which corresponds to the resonant coupling between these two types of states. This precisely corresponds to a form of Feshbach resonance, yet in the frequency domain rather than the magnetic-field domain.
Using this method, we can directly scan the energy spectrum of molecular bound states without synthesizing any molecules. In addition to these bound states, we can also probe the molecular states embedded in the continuum, which are typically very difficult to detect by the conventional methods based on molecular synthesis.
Moreover, by using a far-detuned laser instead of a magnetic field, it enables spatially dependent control over atomic interactions, coupling multiple levels simultaneously, and inducing new Feshbach resonances for those atoms that do not have conventional magnetic resonances. Therefore, we believe that this new resonant mechanism offers new opportunities for controlling atomic and molecular interactions in quantum simulations.
\end{abstract}
\maketitle

The Feshbach resonance \cite{Inouye1998} is a powerful tool in quantum many-body physics. By tuning the magnetic field, two energy levels of atomic or molecular collisional states can cross and couple with each other, leading to the divergence of scattering lengths. This resonant divergence provides a method to adjust the scattering length of ultracold atoms or molecules. Moreover, numerous new quantum phenomena have emerged through the utilization of quantum control techniques based on the Feshbach resonance \cite{RevModPhys.82.1225,RevModPhys.80.885}.
Following this logic, the quantum control community always demands a more versatile tool for adjusting collisional properties, as it may enable the creation of more novel quantum matters \cite{Kraemer2006,Haller2009,Ni2008,LIANG20222550,Liu2018,Yang2022,PhysRevB.106.054310,Park2023,Hu2019,Feng2019,zhang2020,Liu2023,Hanna2007,Zhao2021,Venu2023,Huang2021,Yang2022v2,Zhang2024,Theis2004,Bauer2009,PhysRevLett.115.155301,Chen2023,PhysRevLett.121.163402,Chen2024v2}. The conventional Feshbach resonance \cite{Inouye1998} is based on the adjustment of the magnetic field. However, the fixed resonance magnetic field values and uncontrollable widths impose limitations, especially for magnetic Feshbach resonances with large field values and narrow widths. Besides, it is sometimes challenging to locally address the field or rapidly switch the field direction or magnitude. This is because the magnetic field is created by current coils, which are on a centimeter- or millimeter-scale and have a large inductance that blocks AC oscillations \cite{Hu2019,Feng2019,zhang2020}.

In contrast, laser fields offer greater flexibility, and a variety of techniques have been developed to control Feshbach resonances using lights. In optical Feshbach resonance \cite{Theis2004,PhysRevA.92.022709}, interaction control is allowed by utilizing a laser light to couple the free atomic states with an excited molecular states. The major obstacle of this method is large inelastic loss due to the spontaneous decay of excited states. Optical shifts of a magnetic Feshbach resonance via bound-to-bound transitions \cite{Bauer2009,PhysRevA.88.041601} or a far-detuned laser \cite{PhysRevLett.115.155301} enables reducing such losses but the shifts are usually limited. Apart from laser fields, microwave and radio-frequency (RF) also provide diverse control for cold collisions, such as electric-field-linked resonance \cite{Chen2023,PhysRevLett.121.163402,Chen2024v2} in ultracold dipolar molecular systems. In atomic microwave Feshbach resonance \cite{PhysRevA.81.041603} and RF Feshbach resonance \cite{PhysRevA.81.050701,PhysRevA.94.023619}, an oscillating magnetic field polarized perpendicular to the quantization axis directly couples the incoming atomic states to a bound state. Though these 
approaches are immune to spontaneous decay, the enhancement of pairwise interaction is constrained by small coupling of power-limited microwave or RF. Particularly, a resonance induced by magnetic modulation along the quantization axis \cite{Langmack2015,Hudson2015} can reach greater enhancement because of lower resonance frequency and stronger coupling.

\begin{figure*}[t]
    \centering
    \includegraphics[width=0.9\textwidth]{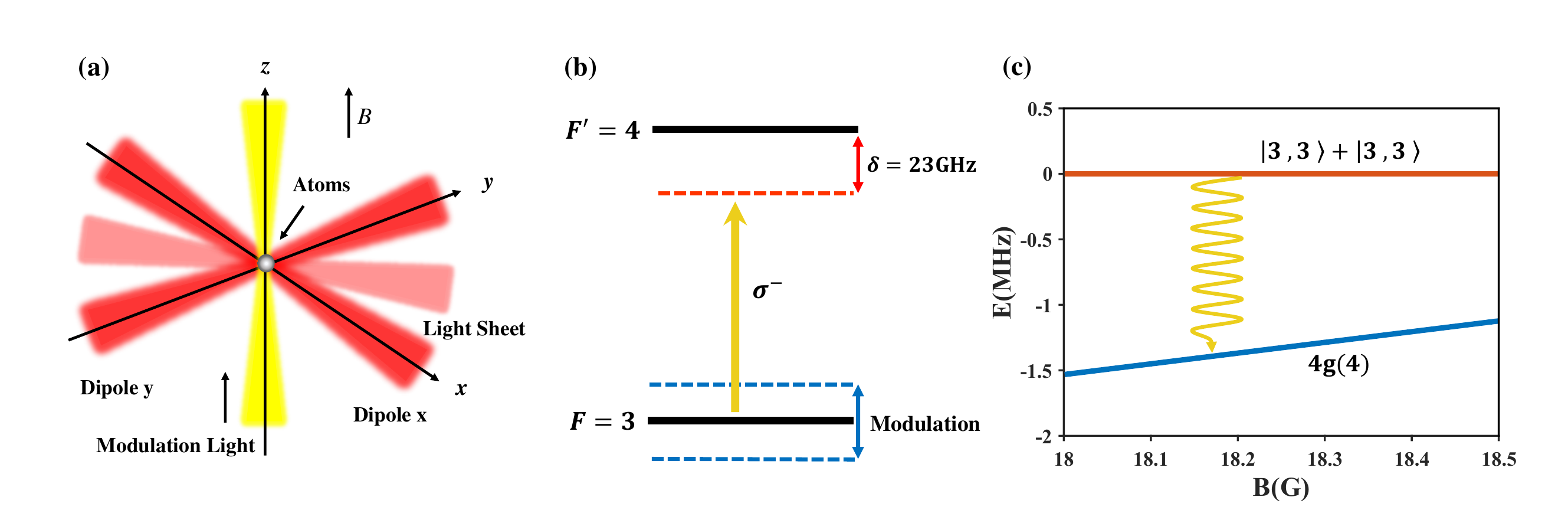}
    \caption{(a) Experimental setup. A Bose-Einstein condensate with approximately $10^5$ cesium atoms, trapped by a light sheet plus a cross-dipole trap in the $x$-$y$ plane, is prepared in the hyperfine state of $|F=3,m_F=3\rangle$ with a magnetic field applied along the $z$ axis. A 23-GHz-detuned light is sent along the $z$ axis with a left-hand circular polarization as depicted in panel (b).
    This light has an intensity modulation with a modulation frequency $\omega$, which oscillates the differential energy shift between atomic collisional states and couples the free-scattering states to the molecular states such as $4g(4)$ in panel (c).}  
    \label{fig1}
\end{figure*}

Following the idea of modulated-magnetic field, we first theoretically extend it to a more generalized scenario. Subsequently, we demonstrate this new phenomenon, the modulation-induced Feshbach resonance, by employing a far-detuned laser instead of a magnetic field.
In previous studies \cite{Langmack2015,Hudson2015}, this new resonance was based on the modulation of the scattering length. According to the Floquet theory, an AC-oscillating scattering length can enhance the DC interaction strength. By tuning the modulation frequency, it is anticipated that the effective scattering length will also resonate with respect to the frequency, which is highly similar to the conventional Feshbach resonance. A prerequisite for this modulated scattering length is the same as that of the conventional resonance: two energy levels must cross at a particular magnetic field \cite{RevModPhys.82.1225,Langmack2015,Hudson2015}, and any change around this specific resonant field will lead to a change in the scattering length.
However, we have discovered that the pre-existence of a two-level crossing or a Feshbach resonance is not a mandatory requirement. The two energy levels of the collisional states do not need to cross with each other. All we need to do is applying a small periodic modulation to one of the energy levels, such that its energy oscillates slightly around its original value. Then, the Floquet theory can induce a new resonance between these two levels, which will directly result in the same behavior as the conventional Feshbach resonance.
This provides a new way to tune the scattering length of atoms that do not have Feshbach resonances in the applicable magnetic fields.

In addition to the theoretical expansions, we conduct experimental demonstrations of this phenomenon using cesium atoms. Here, the energy level modulation is induced by the intensity modulation of a far-detuned laser, which oscillates the energy difference between atomic collisional states. 
Specifically, we illustrate that the free-scattering states of two atoms in the $|F=3,m_F=3\rangle$ state can exhibit frequency-based resonances with the molecular states  $4g(4)$, $4d$ and $6s$ and a near-frequency-independent resonance with $6g(6)$. 
These molecular states are characterized by the quantum numbers $fl(m_f)$ \cite{Mark2007}, where $f$ denotes the resultant of the total atomic angular momentum of the individual atoms, $l$ is the orbital angular momentum of the relative motion, and $m_f$ represents the projection of $f$ along the quantization axis. For states with $m_f = f$, $(m_f)$ is omitted for brevity. This discovery leads to two novel advantages. 
First, we can directly scan out most of the shallow molecular states of cesium atoms, including those states in the continuum. 
Because this measurement is based on the atom loss under a fast modulation of laser light, it is also applicable to those molecular states with a short lifetime. Meanwhile, the molecular states in the continuum may inspire similar new phenomena as the bound state in continuum \cite{Hsu2016,PhysRevA.32.3231}.
Second, in principle, we can simultaneously apply multiple frequencies to modulate the laser intensity, enabling us to bring together many energy levels (more than two) to induce this resonance. Although this might seem conceptually straightforward, it is comparable to the scenarios of two-level and three-level systems in quantum optics. A three-level system is far more complex than a two-level one, as electromagnetic-induced transparency can only occur in a three-level (or multi-level) system. Moreover, numerous quantum applications \cite{Jagannathan2016,Tanji2011,Wenlan2013,Sun2018,Niemietz2021,PhysRevLett.89.067901,Wang20199} are based on this three-level structure.
Therefore, the same reasoning applies to the Feshbach resonance. A recent observation of electromagnetic-induced loss suppression \cite{Jagannathan2016} under the Feshbach resonance and associated atomic collisions serves as supporting evidence, suggesting the potential for tunable interactions involving more levels. In the Supplementary Materials \cite{Supplementary}, we provide more detailed discussions and examples for the interaction control under different scenarios.

\begin{figure*}[t]
    \centering
    \includegraphics[width=1\textwidth]{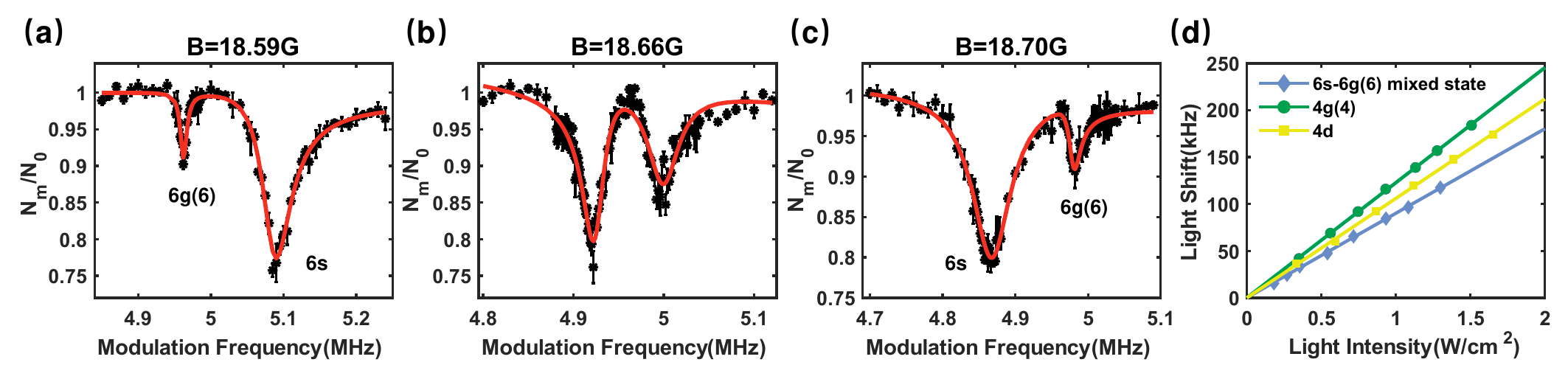}
    \caption{Typical atom loss signals in the frequency domain and dependence of the resonance position shift on the average peak intensity.
    (a)-(c) The relative atom number $N_m/N_0$ versus the modulation frequency $\omega$ after 5 ms exposure to 0.86 W/cm$^2$ (average peak intensity) modulation light at 18.59~G, 18.66~G and 18.70~G, respectively. The resonant peaks are labelled with corresponding molecular states $6g6$ and $6s$ in panels (a) and (c), while these two molecular states are strongly mixed in panel (b). The experimental atom loss signal (dark solid circle) in each panel is fitted with a Fano profile (red solid lines), and all error bars represents one standard deviation of measurements.
    In panel (d), a linear fitting example of the resonance position shift caused by the DC light intensity for $4g(4)$, $4d$ molecular states and one of the $6s$-$6g(6)$ mixed molecular states at respective magnetic field of 17.27~G, 47.36~G and 18.66~G, is presented, which offers a route to compensate the resonance frequency shift.
    }  
    \label{fig2}
\end{figure*}

\begin{figure}[b]
    \centering
    \includegraphics[width=0.45\textwidth]{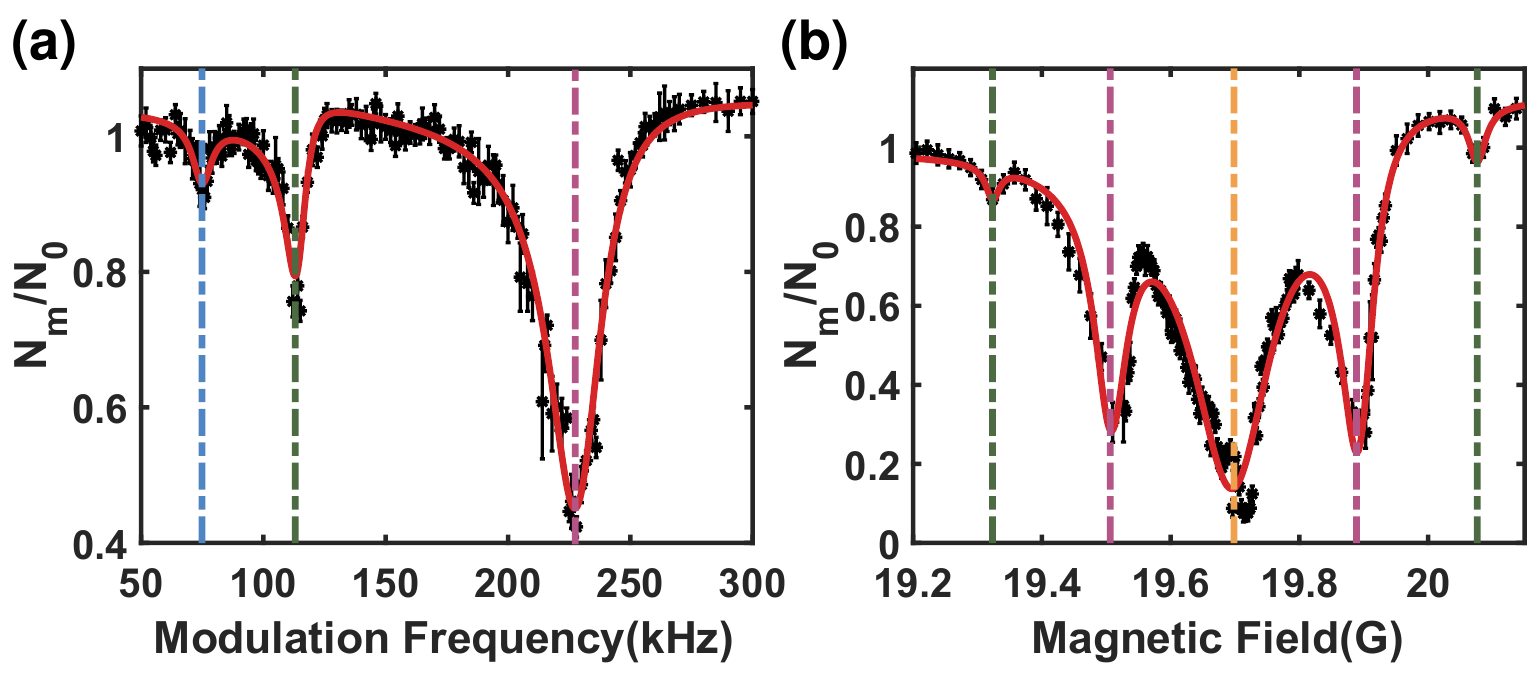}
    \caption{Multiple resonances of the $4g(4)$ molecular state.
    (a) Resonances in the frequency domain at 19.41G.
    (b) Atom-loss feature as a function of the magnetic field under a fixed modulation frequency of 150 kHz. The average peak intensity of applied modulation light is 0.87 W/cm$^2$ in both two panels.
    }  
    \label{fig_extra}
\end{figure}

\begin{figure*}[t]
    \centering
    \includegraphics[width=0.9\textwidth]{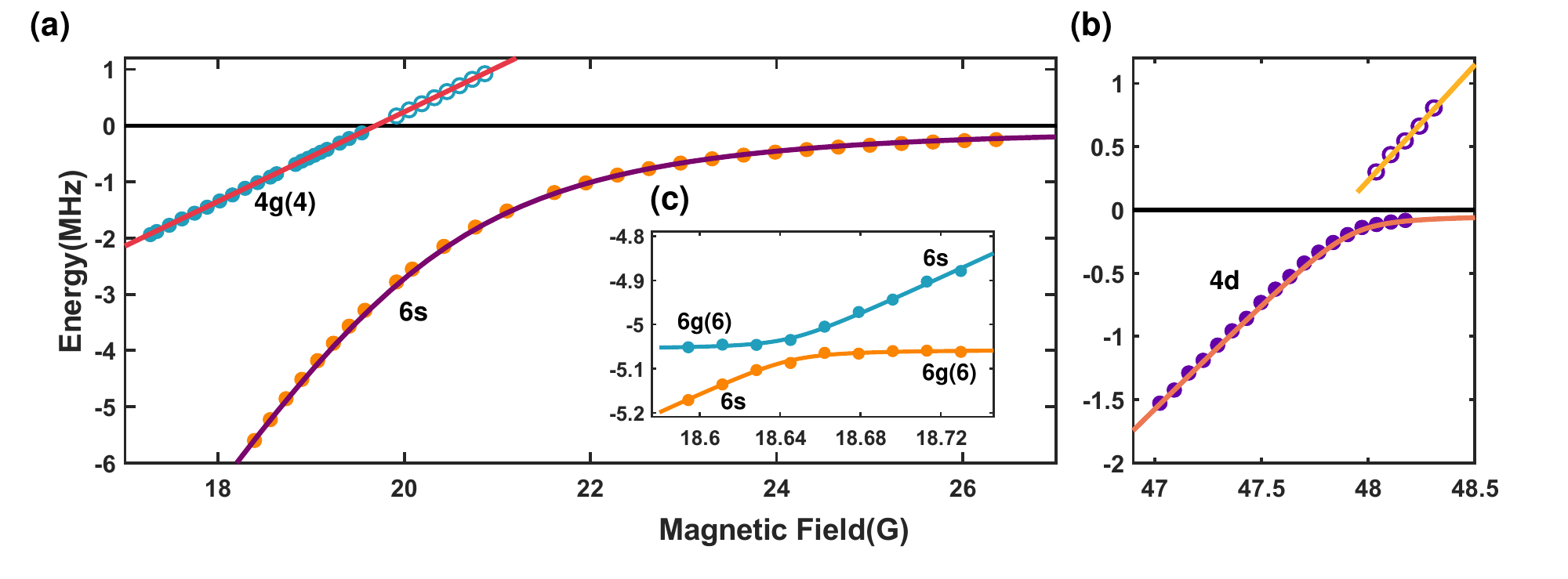}
    \caption{The energy spectrum of cesium molecular states. The filled (unfilled) symbols in the plot represent molecular bound states data (molecular states in the continuum) below (above) the atomic scattering threshold that corresponds to the black solid line along $\omega=0$. The other solid lines are the theoretical models calculated for molecular states $6g(6)$, $6s$,  $4g(4)$ and $4d$. The main plots near 20 G (a) and 48 G (b) depict the energy of $4g(4)$, $6s$ and $4d$, exhibiting obvious dependence on magnetic field. In the inset (c), an avoid crossing between the $6s$ state and the near-frequency-independent $6g(6)$ state is shown. This feature is not plotted in (a) as it's a narrow span near 18.66 G.
    The $4g(4)$ and $4d$ data are obtained at a average peak intensity of 0.86 W/cm$^2$ and compared with theoretical lines shifted from their original Feshbach resonance points to the optically shifted ones measured in experiments, while the $6s$ and $6g6$ data are obtained by varying the average peak intensity and compensating the DC component of the light shift by linear fitting.
    }  
    \label{fig4}
\end{figure*}

Now we demonstrate how we extend the idea of the modulated magnetic Feshbach resonance using a two-level model. This approach retains the essential physics while neglecting microscopic details of the scattering interactions. Assume we have two states,  $|\alpha\rangle$ and $|\beta\rangle$, with corresponding uncoupled energies $\hbar\omega_\alpha$ and $\hbar\omega_\beta$, where $\hbar$ is the reduced Planck constant, and define $\omega_b=\omega_\alpha - \omega_\beta$ to denote the energy difference.
There exists a non-zero off-diagonal coupling term $\hbar\Omega/2$ between these two states due to non-zero wave function overlap.
Without loss of generality, we consider that the energy of $|\beta\rangle$ is periodically modulated in the form of $\hbar\omega_\beta+\hbar A\cos(\omega t)$, where $A$ is the modulation amplitude and $\omega$ is the modulation frequency. This modulation term $\hbar A\cos(\omega t)$ can stem from the differential shift between these two states caused by an intensity-modulated light. We write the Hamiltonian $H_0(t)$ in matrix form as:
\begin{equation}
H_0(t)=\hbar
\left(
\begin{array}{cc}
\omega_
\alpha  & \Omega/2     \\
\Omega/2  & \omega_\beta+A\cos(\omega t)\end{array}
\right).
\end{equation}

To transform this original Hamiltonian into a rotating frame, a unitary transformation $U(t)$ is applied. The new Hamiltonian under transformation is $H_1(t)=U (t) H_0(t) U^\dagger (t)-i\hbar U(t) \partial U^\dagger(t)/\partial t$. When choose $U(t)$ in the diagonal form $U(t)=\textrm{Diag}\{1,e^{i\left[m\omega t+\frac{A}{\omega}\sin(\omega t)\right]}   \}$, where $m$ is an integer, $H_1(t)$ becomes:
\begin{equation}
H_1(t)=\hbar
\left(
\begin{array}{cc}
\omega_
\alpha  & \frac{\Omega}{2} e^{-i\left[m \omega t+\frac{A}{\omega}\sin(\omega t)\right]}    \\
\frac{\Omega}{2} e^{i\left[m \omega t+\frac{A}{\omega}\sin(\omega t)\right]}  & \omega_\alpha-\omega_b-m \omega\end{array}
\right).
\end{equation}

By using the expansion $e^{i\frac{A}{\omega}\sin(\omega t)}=\sum_n J_n (A/\omega)e^{in\omega t}$, where $J_n (x)$ is the $n$-th order Bessel function of the first kind, we rewrite $H_1$ as:
\begin{eqnarray}
& &\frac{H_1(t)}{\hbar}=\omega_\alpha+ \nonumber \\
& &
\left(
\begin{array}{cc}
0  &\sum_n \frac{\Omega}{2}J_n(\frac{A}{\omega}) e^{-i(m+n)\omega t}   \\
\sum_n\frac{\Omega}{2}J_n(\frac{A}{\omega}) e^{i(m+n)\omega t}  & -\omega_b-m \omega\end{array}
\right).
\end{eqnarray}

The first term $\omega_\alpha$ is a constant and can be ignored. The second term describes the interaction between the two states. By retaining the resonant term under the condition $n=-m$, the interaction Hamiltonian is simplified to:
\begin{equation}
\frac{H_1(t)}{\hbar}-\omega_\alpha=
\left(
\begin{array}{cc}
0  & \frac{\Omega}{2} (-1)^m J_{m}(\frac{A}{\omega})    \\
\frac{\Omega}{2} (-1)^m J_{m}(\frac{A}{\omega})  & -\omega_b-m\omega\end{array}
\right).
\end{equation}

Therefore, when $m \omega$ approaches $- \omega_b=\omega_\beta-\omega_\alpha$, the system becomes a resonant-coupled two-level system. Detailed derivations of its effects on scattering properties and scattering lengths are provided in \cite{Supplementary}. Here, we present the conclusion that due to the resonant coupling from $|\alpha\rangle$ to $|\beta\rangle$, the effective scattering length $a_s$ of the incoming state $|\alpha\rangle$ has the form:

\begin{equation}
a_s=a_{\textrm{BK}}\left(1-\frac{\Delta_m}{-m\omega-\omega_0}\right).
\label{eq5}
\end{equation}

Here $a_{\textrm{BK}}$ is the background scattering length, $\Delta_m$ is the width of the resonance and $\omega_0$ is close to $ \omega_b$. 
It exactly corresponds to the form of Feshbach resonance, but with the magnetic field replaced by an integer multiple of the modulation frequency $\omega$, which we use the term, modulation-induced Feshbach resonance, to describe it. 
More discussions about this resonance mechanism are included in \cite{Supplementary}. Comparing to the previous study by magnetic modulation, here the two energy levels $|\alpha\rangle$ and $|\beta\rangle$ do not need any crossing or preexisted Feshbach resonance, as it only need tiny level modulation.

We now turn to the experimental implementation. We prepare a Bose-Einstein condensate consisting of approximately $10^5$ cesium atoms, with all atoms initialized in the $6S$ hyperfine ground state $|F=3,m_F=3\rangle$ \cite{Supplementary}.
A magnetic field $B$ is applied along the $z$-axis (as depicted in Fig.~\ref{fig1}a) and serves as the quantization axis.
Subsequently, an intensity-modulated light with a left-hand circularly polarization and a 23 GHz red-detuning relative to the cesium $D_2$ transition of $F=3\rightarrow F'=4$ (Fig.~\ref{fig1}b), is directed along the $z$-axis towards the atoms, inducing light shifts on free-scattering states and molecular states \cite{Supplementary}. The average peak intensity is varied from 0.2 W/cm$^2$ to 1.7 W/cm$^2$ in our experiments, while the modulation depth is fixed to be $86\%$.
Then we sweep the modulation frequencies $\omega$ with the aim of coupling the molecular states (Fig.~\ref{fig1}c). After irradiating the atoms with the light for 1-20 ms, we switch off the light and measure the remaining atom number~\cite{Liu2023}. The modulation-induced Feshbach resonances are detected by atom loss signals and their positions are extracted from a fit.

In Fig.~\ref{fig2}a-c, we present data obtained at different magnetic fields (18.59, 18.66, and 18.70~G). All panels exhibit significant atom loss, where $N_m/N_0$ is the relative atom number, $N_m$ is the remaining atom number after modulation, and $N_0$ is the atom number far away from resonance positions. These loss signals correspond to the molecular states $6s$ and $6g(6)$. Especially, the energy level of $6g(6)$ is nearly parallel to the incoming scattering state (both atoms in $|F=3,m_F=3\rangle$) as a function of the magnetic field, and this state does not have a direct energy crossing with the scattering state at low magnetic field regime~\cite{2012model}. 
These experimental results support our hypothesis that modulation can induce a resonance between two states without the pre-existence of a conventional Feshbach resonance.
Furthermore, other locations correspond to the molecular states $4g(4)$ and $4d$ are also observed, and their resonant frequencies exhibit obvious dependence on the magnetic field like the molecular state $6s$. 
Notably, compared to the original resonance frequencies, the measured atom loss peaks are all red-shifted due to the DC component of the light shift. This shift can be determined and compensated by linearly fitting between the atom loss peak position and the average peak intensity, as shown in Fig.~\ref{fig2}d.
The atom-loss characteristics as a function of frequency precisely mirror the form of the conventional Feshbach resonance, where atom-loss features are measured as a function of the magnetic field. This type of resonance in the frequency domain is what we term the modulation-induced Feshbach resonance.

In most cases, there is only one resonance peak for each molecular state, which corresponds to the first-order modulation resonance for $|m| =  1$  in Eq. (\ref{eq5}) ($m = -1$ for molecular states below the atomic scattering threshold, $m = 1$ for molecular states above the atomic scattering threshold). However, when the modulation amplitude $A$ is comparable to the $|m|>1$ multiple of the energy difference $\omega_b$ between the free-scattering state and the molecular state, the higher-order modulation resonances emerge. For the $4g(4)$ molecular state at magnetic fields where the binding energy is relatively small, we clearly observe these higher-order resonances. As exhibited in Fig.~\ref{fig_extra}a, the dominant loss peak at $f_1=228.7$~kHz (purple dash-dotted line) corresponds to the first-order modulation resonance at 19.41~G. The peaks at frequencies $f_2=114.3$~kHz$ \approx f_1/2$ and $f_3=75.0$~kHz$ \approx f_1/3$ indicate the higher-order modulation resonances (green and blue dash-dotted lines).
Moreover, these modulation effects actually also modify the atom-loss feature as a function of the magnetic field, which can be understood by fixing the modulation frequency $\omega$ and varying $\omega_0$ with the magnetic field in Eq. (\ref{eq5}), and is experimentally shown in Fig.~\ref{fig_extra}b by sweeping the magnetic field and measuring the relative atom number $N_m/N_0$ at a fixed modulation frequency of 150 kHz. Surprisingly, the Feshbach resonance is significantly shifted from 19.84 G to 19.70 G (orange dash-dotted line) due to the light shift  \cite{Supplementary}. The peaks at 19.51 G and 19.90 G (purple dash-dotted line) are attributed to the first-order modulation resonances, while those at 19.32 G and 20.08 G (green dash-dotted line) arise from the second-order modulation resonances. The resonance features appearing at the left (right) side of the shifted Feshbach resonance arise from coupling between free-scattering states and bound molecular states (molecular states embedded in the continuum).

In Fig.~\ref{fig4}, the measured resonant frequencies as a function of the magnetic field as well as their theoretical values \cite{2009fitting,2012model} are plotted.
The vertical axis of the modulation frequency $\omega$ is inverted for the bound states whose energies are below those of the free-scattering states. In addition to the bound states, we also identify the molecular states embedded in the continuum with unfilled symbols in the plot to represent the corresponding measurement results. 
The measured resonance frequencies exhibit an accuracy of $\sim$ 10 kHz, limited primarily by magnetic field fluctuations \cite{Supplementary}. This is akin to the previous studies using magnetic moment spectroscopy \cite{Mark2007} and microwave spectroscopy \cite{Mark2007,PhysRevLett.97.120402}. A further comparison of these methods is presented in \cite{Supplementary}.

Beyond the overall view of the energy spectrum, there are also intriguing detailed features observable in this plot. In Fig.~\ref{fig4}c, we zoom in on the crossing region of the $6g(6)$ and $6s$ states. This region exhibits the characteristic features of a Landau-Zener crossing, where two energy levels intersect and a coupling term between them creates an energy gap.
By fitting the data to the Landau-Zener model where the energies are decided by $ E_{\pm} = \left[{E_i+E_j\pm\sqrt{(E_i-E_j)^2+V^2_{ij}}}\right]/{2}$, we determine that the Rabi frequency $V_{ij}/h$ of this coupling between these two states is 25~kHz. Our experimental data reveal the $6g(6)$, $6s$, $4d$, $4g(4)$ molecular states and exhibit excellent agreement with theoretical calculations, demonstrating the effectiveness of our method for probing deeper and frequency-independent molecular states compared to the magnetic modulation \cite{2009fitting,2012model,PhysRevLett.95.190404}.

In conclusion, we have experimentally observed a novel resonant mechanism, the modulation-induced Feshbach resonance, through the utilization of an intensity-modulated light. This new resonance shares a similar form with the conventional Feshbach resonance, yet it occurs in the frequency domain rather than relying on a tunable magnetic field. It enables us to directly map out the energy spectrum of shallow molecular states, including those embedded in the continuum. Given the highly versatile tunability of lasers, in principle, we can bring together more than two energy levels to induce the resonance, which we anticipate will surpass the capabilities of the conventional Feshbach resonance. We look forward to the further development of future experimental investigations and associated applications based on a three-level resonance.

This work is supported by the National Key Research and Development Program of China (2021YFA0718303, 2021YFA1400904, and 2023YFA1406702) and the National Natural Science Foundation of China (92165203 and 92476110).


\begin{thebibliography}{49}%
\makeatletter
\providecommand \@ifxundefined [1]{%
 \@ifx{#1\undefined}
}%
\providecommand \@ifnum [1]{%
 \ifnum #1\expandafter \@firstoftwo
 \else \expandafter \@secondoftwo
 \fi
}%
\providecommand \@ifx [1]{%
 \ifx #1\expandafter \@firstoftwo
 \else \expandafter \@secondoftwo
 \fi
}%
\providecommand \natexlab [1]{#1}%
\providecommand \enquote  [1]{``#1''}%
\providecommand \bibnamefont  [1]{#1}%
\providecommand \bibfnamefont [1]{#1}%
\providecommand \citenamefont [1]{#1}%
\providecommand \href@noop [0]{\@secondoftwo}%
\providecommand \href [0]{\begingroup \@sanitize@url \@href}%
\providecommand \@href[1]{\@@startlink{#1}\@@href}%
\providecommand \@@href[1]{\endgroup#1\@@endlink}%
\providecommand \@sanitize@url [0]{\catcode `\\12\catcode `\$12\catcode
  `\&12\catcode `\#12\catcode `\^12\catcode `\_12\catcode `\%12\relax}%
\providecommand \@@startlink[1]{}%
\providecommand \@@endlink[0]{}%
\providecommand \url  [0]{\begingroup\@sanitize@url \@url }%
\providecommand \@url [1]{\endgroup\@href {#1}{\urlprefix }}%
\providecommand \urlprefix  [0]{URL }%
\providecommand \Eprint [0]{\href }%
\providecommand \doibase [0]{http://dx.doi.org/}%
\providecommand \selectlanguage [0]{\@gobble}%
\providecommand \bibinfo  [0]{\@secondoftwo}%
\providecommand \bibfield  [0]{\@secondoftwo}%
\providecommand \translation [1]{[#1]}%
\providecommand \BibitemOpen [0]{}%
\providecommand \bibitemStop [0]{}%
\providecommand \bibitemNoStop [0]{.\EOS\space}%
\providecommand \EOS [0]{\spacefactor3000\relax}%
\providecommand \BibitemShut  [1]{\csname bibitem#1\endcsname}%
\let\auto@bib@innerbib\@empty
%</preamble>
\bibitem [{\citenamefont {Inouye}\ \emph {et~al.}(1998)\citenamefont {Inouye}, \citenamefont {Andrews}, \citenamefont {Stenger}, \citenamefont {Miesner}, \citenamefont {Stamper-Kurn},\ and\ \citenamefont {Ketterle}}]{Inouye1998}%
  \BibitemOpen
  \bibfield  {author} {\bibinfo {author} {\bibfnamefont {S.}~\bibnamefont {Inouye}}, \bibinfo {author} {\bibfnamefont {M.~R.}\ \bibnamefont {Andrews}}, \bibinfo {author} {\bibfnamefont {J.}~\bibnamefont {Stenger}}, \bibinfo {author} {\bibfnamefont {H.~J.}\ \bibnamefont {Miesner}}, \bibinfo {author} {\bibfnamefont {D.~M.}\ \bibnamefont {Stamper-Kurn}},\ and\ \bibinfo {author} {\bibfnamefont {W.}~\bibnamefont {Ketterle}},\ }\bibfield  {title} {\bibinfo {title} {{Observation of Feshbach resonances in a Bose--Einstein condensate}},\ }\href {https://doi.org/10.1038/32354} {\bibfield  {journal} {\bibinfo  {journal} {Nature}\ }\textbf {\bibinfo {volume} {392}},\ \bibinfo {pages} {151} (\bibinfo {year} {1998})}\BibitemShut {NoStop}%
\bibitem [{\citenamefont {Chin}\ \emph {et~al.}(2010)\citenamefont {Chin}, \citenamefont {Grimm}, \citenamefont {Julienne},\ and\ \citenamefont {Tiesinga}}]{RevModPhys.82.1225}%
  \BibitemOpen
  \bibfield  {author} {\bibinfo {author} {\bibfnamefont {C.}~\bibnamefont {Chin}}, \bibinfo {author} {\bibfnamefont {R.}~\bibnamefont {Grimm}}, \bibinfo {author} {\bibfnamefont {P.}~\bibnamefont {Julienne}},\ and\ \bibinfo {author} {\bibfnamefont {E.}~\bibnamefont {Tiesinga}},\ }\bibfield  {title} {\bibinfo {title} {Feshbach resonances in ultracold gases},\ }\href {https://doi.org/10.1103/RevModPhys.82.1225} {\bibfield  {journal} {\bibinfo  {journal} {{Rev. Mod. Phys.}}\ }\textbf {\bibinfo {volume} {82}},\ \bibinfo {pages} {1225} (\bibinfo {year} {2010})}\BibitemShut {NoStop}%
\bibitem [{\citenamefont {Bloch}\ \emph {et~al.}(2008)\citenamefont {Bloch}, \citenamefont {Dalibard},\ and\ \citenamefont {Zwerger}}]{RevModPhys.80.885}%
  \BibitemOpen
  \bibfield  {author} {\bibinfo {author} {\bibfnamefont {I.}~\bibnamefont {Bloch}}, \bibinfo {author} {\bibfnamefont {J.}~\bibnamefont {Dalibard}},\ and\ \bibinfo {author} {\bibfnamefont {W.}~\bibnamefont {Zwerger}},\ }\bibfield  {title} {\bibinfo {title} {Many-body physics with ultracold gases},\ }\href {https://doi.org/10.1103/RevModPhys.80.885} {\bibfield  {journal} {\bibinfo  {journal} {{Rev. Mod. Phys.}}\ }\textbf {\bibinfo {volume} {80}},\ \bibinfo {pages} {885} (\bibinfo {year} {2008})}\BibitemShut {NoStop}%
\bibitem [{\citenamefont {Kraemer}\ \emph {et~al.}(2006)\citenamefont {Kraemer}, \citenamefont {Mark}, \citenamefont {Waldburger}, \citenamefont {Danzl}, \citenamefont {Chin}, \citenamefont {Engeser}, \citenamefont {Lange}, \citenamefont {Pilch}, \citenamefont {Jaakkola}, \citenamefont {N{\"a}gerl},\ and\ \citenamefont {Grimm}}]{Kraemer2006}%
  \BibitemOpen
  \bibfield  {author} {\bibinfo {author} {\bibfnamefont {T.}~\bibnamefont {Kraemer}}, \bibinfo {author} {\bibfnamefont {M.}~\bibnamefont {Mark}}, \bibinfo {author} {\bibfnamefont {P.}~\bibnamefont {Waldburger}}, \bibinfo {author} {\bibfnamefont {J.~G.}\ \bibnamefont {Danzl}}, \bibinfo {author} {\bibfnamefont {C.}~\bibnamefont {Chin}}, \bibinfo {author} {\bibfnamefont {B.}~\bibnamefont {Engeser}}, \bibinfo {author} {\bibfnamefont {A.~D.}\ \bibnamefont {Lange}}, \bibinfo {author} {\bibfnamefont {K.}~\bibnamefont {Pilch}}, \bibinfo {author} {\bibfnamefont {A.}~\bibnamefont {Jaakkola}}, \bibinfo {author} {\bibfnamefont {H.~C.}\ \bibnamefont {N{\"a}gerl}},\ and\ \bibinfo {author} {\bibfnamefont {R.}~\bibnamefont {Grimm}},\ }\bibfield  {title} {\bibinfo {title} {Evidence for {E}fimov quantum states in an ultracold gas of caesium atoms},\ }\href {https://doi.org/10.1038/nature04626} {\bibfield  {journal} {\bibinfo  {journal} {Nature}\ }\textbf {\bibinfo {volume} {440}},\ \bibinfo {pages} {315} (\bibinfo {year}
  {2006})}\BibitemShut {NoStop}%
\bibitem [{\citenamefont {Haller}\ \emph {et~al.}(2009)\citenamefont {Haller}, \citenamefont {Gustavsson}, \citenamefont {Mark}, \citenamefont {Danzl}, \citenamefont {Hart}, \citenamefont {Pupillo},\ and\ \citenamefont {Nägerl}}]{Haller2009}%
  \BibitemOpen
  \bibfield  {author} {\bibinfo {author} {\bibfnamefont {E.}~\bibnamefont {Haller}}, \bibinfo {author} {\bibfnamefont {M.}~\bibnamefont {Gustavsson}}, \bibinfo {author} {\bibfnamefont {M.~J.}\ \bibnamefont {Mark}}, \bibinfo {author} {\bibfnamefont {J.~G.}\ \bibnamefont {Danzl}}, \bibinfo {author} {\bibfnamefont {R.}~\bibnamefont {Hart}}, \bibinfo {author} {\bibfnamefont {G.}~\bibnamefont {Pupillo}},\ and\ \bibinfo {author} {\bibfnamefont {H.-C.}\ \bibnamefont {Nägerl}},\ }\bibfield  {title} {\bibinfo {title} {Realization of an excited, strongly correlated quantum gas phase},\ }\href {https://doi.org/10.1126/science.1175850} {\bibfield  {journal} {\bibinfo  {journal} {Science}\ }\textbf {\bibinfo {volume} {325}},\ \bibinfo {pages} {1224} (\bibinfo {year} {2009})}\BibitemShut {NoStop}%
\bibitem [{\citenamefont {Ni}\ \emph {et~al.}(2008)\citenamefont {Ni}, \citenamefont {Ospelkaus}, \citenamefont {de~Miranda}, \citenamefont {Pe'er}, \citenamefont {Neyenhuis}, \citenamefont {Zirbel}, \citenamefont {Kotochigova}, \citenamefont {Julienne}, \citenamefont {Jin},\ and\ \citenamefont {Ye}}]{Ni2008}%
  \BibitemOpen
  \bibfield  {author} {\bibinfo {author} {\bibfnamefont {K.-K.}\ \bibnamefont {Ni}}, \bibinfo {author} {\bibfnamefont {S.}~\bibnamefont {Ospelkaus}}, \bibinfo {author} {\bibfnamefont {M.~H.~G.}\ \bibnamefont {de~Miranda}}, \bibinfo {author} {\bibfnamefont {A.}~\bibnamefont {Pe'er}}, \bibinfo {author} {\bibfnamefont {B.}~\bibnamefont {Neyenhuis}}, \bibinfo {author} {\bibfnamefont {J.~J.}\ \bibnamefont {Zirbel}}, \bibinfo {author} {\bibfnamefont {S.}~\bibnamefont {Kotochigova}}, \bibinfo {author} {\bibfnamefont {P.~S.}\ \bibnamefont {Julienne}}, \bibinfo {author} {\bibfnamefont {D.~S.}\ \bibnamefont {Jin}},\ and\ \bibinfo {author} {\bibfnamefont {J.}~\bibnamefont {Ye}},\ }\bibfield  {title} {\bibinfo {title} {A high phase-space-density gas of polar molecules},\ }\href {https://doi.org/10.1126/science.1163861} {\bibfield  {journal} {\bibinfo  {journal} {Science}\ }\textbf {\bibinfo {volume} {322}},\ \bibinfo {pages} {231} (\bibinfo {year} {2008})}\BibitemShut {NoStop}%
\bibitem [{\citenamefont {Liang}\ \emph {et~al.}(2022)\citenamefont {Liang}, \citenamefont {Zheng}, \citenamefont {Yao}, \citenamefont {Zheng}, \citenamefont {Yao}, \citenamefont {Zhou}, \citenamefont {Huang}, \citenamefont {Zhang}, \citenamefont {Ye}, \citenamefont {Zhou}, \citenamefont {Chen}, \citenamefont {Chen}, \citenamefont {Zhai},\ and\ \citenamefont {Hu}}]{LIANG20222550}%
  \BibitemOpen
  \bibfield  {author} {\bibinfo {author} {\bibfnamefont {L.}~\bibnamefont {Liang}}, \bibinfo {author} {\bibfnamefont {W.}~\bibnamefont {Zheng}}, \bibinfo {author} {\bibfnamefont {R.}~\bibnamefont {Yao}}, \bibinfo {author} {\bibfnamefont {Q.}~\bibnamefont {Zheng}}, \bibinfo {author} {\bibfnamefont {Z.}~\bibnamefont {Yao}}, \bibinfo {author} {\bibfnamefont {T.-G.}\ \bibnamefont {Zhou}}, \bibinfo {author} {\bibfnamefont {Q.}~\bibnamefont {Huang}}, \bibinfo {author} {\bibfnamefont {Z.}~\bibnamefont {Zhang}}, \bibinfo {author} {\bibfnamefont {J.}~\bibnamefont {Ye}}, \bibinfo {author} {\bibfnamefont {X.}~\bibnamefont {Zhou}}, \bibinfo {author} {\bibfnamefont {X.}~\bibnamefont {Chen}}, \bibinfo {author} {\bibfnamefont {W.}~\bibnamefont {Chen}}, \bibinfo {author} {\bibfnamefont {H.}~\bibnamefont {Zhai}},\ and\ \bibinfo {author} {\bibfnamefont {J.}~\bibnamefont {Hu}},\ }\bibfield  {title} {\bibinfo {title} {Probing quantum many-body correlations by universal ramping dynamics},\ }\href
  {https://doi.org/https://doi.org/10.1016/j.scib.2022.12.005} {\bibfield  {journal} {\bibinfo  {journal} {{Science Bulletin}}\ }\textbf {\bibinfo {volume} {67}},\ \bibinfo {pages} {2550} (\bibinfo {year} {2022})}\BibitemShut {NoStop}%
\bibitem [{\citenamefont {Liu}\ \emph {et~al.}(2018)\citenamefont {Liu}, \citenamefont {Hood}, \citenamefont {Yu}, \citenamefont {Zhang}, \citenamefont {Hutzler}, \citenamefont {Rosenband},\ and\ \citenamefont {Ni}}]{Liu2018}%
  \BibitemOpen
  \bibfield  {author} {\bibinfo {author} {\bibfnamefont {L.~R.}\ \bibnamefont {Liu}}, \bibinfo {author} {\bibfnamefont {J.~D.}\ \bibnamefont {Hood}}, \bibinfo {author} {\bibfnamefont {Y.}~\bibnamefont {Yu}}, \bibinfo {author} {\bibfnamefont {J.~T.}\ \bibnamefont {Zhang}}, \bibinfo {author} {\bibfnamefont {N.~R.}\ \bibnamefont {Hutzler}}, \bibinfo {author} {\bibfnamefont {T.}~\bibnamefont {Rosenband}},\ and\ \bibinfo {author} {\bibfnamefont {K.-K.}\ \bibnamefont {Ni}},\ }\bibfield  {title} {\bibinfo {title} {Building one molecule from a reservoir of two atoms},\ }\href {https://doi.org/10.1126/science.aar7797} {\bibfield  {journal} {\bibinfo  {journal} {Science}\ }\textbf {\bibinfo {volume} {360}},\ \bibinfo {pages} {900} (\bibinfo {year} {2018})}\BibitemShut {NoStop}%
\bibitem [{\citenamefont {Yang}\ \emph {et~al.}(2022{\natexlab{a}})\citenamefont {Yang}, \citenamefont {Wang}, \citenamefont {Su}, \citenamefont {Cao}, \citenamefont {Zhang}, \citenamefont {Rui}, \citenamefont {Zhao}, \citenamefont {Bai},\ and\ \citenamefont {Pan}}]{Yang2022}%
  \BibitemOpen
  \bibfield  {author} {\bibinfo {author} {\bibfnamefont {H.}~\bibnamefont {Yang}}, \bibinfo {author} {\bibfnamefont {X.-Y.}\ \bibnamefont {Wang}}, \bibinfo {author} {\bibfnamefont {Z.}~\bibnamefont {Su}}, \bibinfo {author} {\bibfnamefont {J.}~\bibnamefont {Cao}}, \bibinfo {author} {\bibfnamefont {D.-C.}\ \bibnamefont {Zhang}}, \bibinfo {author} {\bibfnamefont {J.}~\bibnamefont {Rui}}, \bibinfo {author} {\bibfnamefont {B.}~\bibnamefont {Zhao}}, \bibinfo {author} {\bibfnamefont {C.-L.}\ \bibnamefont {Bai}},\ and\ \bibinfo {author} {\bibfnamefont {J.-W.}\ \bibnamefont {Pan}},\ }\bibfield  {title} {\bibinfo {title} {Evidence for the association of triatomic molecules in ultracold $^{23}${Na}$^{40}${K} + $^{40}${K} mixtures},\ }\href {https://doi.org/10.1038/s41586-021-04297-2} {\bibfield  {journal} {\bibinfo  {journal} {Nature}\ }\textbf {\bibinfo {volume} {602}},\ \bibinfo {pages} {229} (\bibinfo {year} {2022}{\natexlab{a}})}\BibitemShut {NoStop}%
\bibitem [{\citenamefont {Miao}\ \emph {et~al.}(2022)\citenamefont {Miao}, \citenamefont {Zhang}, \citenamefont {Zhao}, \citenamefont {Zhao}, \citenamefont {Wang},\ and\ \citenamefont {Hu}}]{PhysRevB.106.054310}%
  \BibitemOpen
  \bibfield  {author} {\bibinfo {author} {\bibfnamefont {S.}~\bibnamefont {Miao}}, \bibinfo {author} {\bibfnamefont {Z.}~\bibnamefont {Zhang}}, \bibinfo {author} {\bibfnamefont {Y.}~\bibnamefont {Zhao}}, \bibinfo {author} {\bibfnamefont {Z.}~\bibnamefont {Zhao}}, \bibinfo {author} {\bibfnamefont {H.}~\bibnamefont {Wang}},\ and\ \bibinfo {author} {\bibfnamefont {J.}~\bibnamefont {Hu}},\ }\bibfield  {title} {\bibinfo {title} {{Bosonic fractional quantum Hall conductance in shaken honeycomb optical lattices without flat bands}},\ }\href {https://doi.org/10.1103/PhysRevB.106.054310} {\bibfield  {journal} {\bibinfo  {journal} {{Phys. Rev. B}}\ }\textbf {\bibinfo {volume} {106}},\ \bibinfo {pages} {054310} (\bibinfo {year} {2022})}\BibitemShut {NoStop}%
\bibitem [{\citenamefont {Park}\ \emph {et~al.}(2023)\citenamefont {Park}, \citenamefont {Lu}, \citenamefont {Jamison}, \citenamefont {Tscherbul},\ and\ \citenamefont {Ketterle}}]{Park2023}%
  \BibitemOpen
  \bibfield  {author} {\bibinfo {author} {\bibfnamefont {J.~J.}\ \bibnamefont {Park}}, \bibinfo {author} {\bibfnamefont {Y.-K.}\ \bibnamefont {Lu}}, \bibinfo {author} {\bibfnamefont {A.~O.}\ \bibnamefont {Jamison}}, \bibinfo {author} {\bibfnamefont {T.~V.}\ \bibnamefont {Tscherbul}},\ and\ \bibinfo {author} {\bibfnamefont {W.}~\bibnamefont {Ketterle}},\ }\bibfield  {title} {\bibinfo {title} {{A Feshbach resonance in collisions between triplet ground-state molecules}},\ }\href {https://doi.org/10.1038/s41586-022-05635-8} {\bibfield  {journal} {\bibinfo  {journal} {Nature}\ }\textbf {\bibinfo {volume} {614}},\ \bibinfo {pages} {54} (\bibinfo {year} {2023})}\BibitemShut {NoStop}%
\bibitem [{\citenamefont {Hu}\ \emph {et~al.}(2019)\citenamefont {Hu}, \citenamefont {Feng}, \citenamefont {Zhang},\ and\ \citenamefont {Chin}}]{Hu2019}%
  \BibitemOpen
  \bibfield  {author} {\bibinfo {author} {\bibfnamefont {J.}~\bibnamefont {Hu}}, \bibinfo {author} {\bibfnamefont {L.}~\bibnamefont {Feng}}, \bibinfo {author} {\bibfnamefont {Z.}~\bibnamefont {Zhang}},\ and\ \bibinfo {author} {\bibfnamefont {C.}~\bibnamefont {Chin}},\ }\bibfield  {title} {\bibinfo {title} {{Quantum simulation of Unruh radiation}},\ }\href {https://doi.org/10.1038/s41567-019-0537-1} {\bibfield  {journal} {\bibinfo  {journal} {{Nature Physics}}\ }\textbf {\bibinfo {volume} {15}},\ \bibinfo {pages} {785} (\bibinfo {year} {2019})}\BibitemShut {NoStop}%
\bibitem [{\citenamefont {Feng}\ \emph {et~al.}(2019)\citenamefont {Feng}, \citenamefont {Hu}, \citenamefont {Clark},\ and\ \citenamefont {Chin}}]{Feng2019}%
  \BibitemOpen
  \bibfield  {author} {\bibinfo {author} {\bibfnamefont {L.}~\bibnamefont {Feng}}, \bibinfo {author} {\bibfnamefont {J.}~\bibnamefont {Hu}}, \bibinfo {author} {\bibfnamefont {L.~W.}\ \bibnamefont {Clark}},\ and\ \bibinfo {author} {\bibfnamefont {C.}~\bibnamefont {Chin}},\ }\bibfield  {title} {\bibinfo {title} {{Correlations in high harmonic generation of matter-wave jets revealed by pattern recognition}},\ }\href {https://doi.org/10.1126/science.aat5008} {\bibfield  {journal} {\bibinfo  {journal} {{Science}}\ }\textbf {\bibinfo {volume} {524}},\ \bibinfo {pages} {521} (\bibinfo {year} {2019})}\BibitemShut {NoStop}%
\bibitem [{\citenamefont {Zhang}\ \emph {et~al.}(2020)\citenamefont {Zhang}, \citenamefont {Yao}, \citenamefont {Feng}, \citenamefont {Hu},\ and\ \citenamefont {Chin}}]{zhang2020}%
  \BibitemOpen
  \bibfield  {author} {\bibinfo {author} {\bibfnamefont {Z.}~\bibnamefont {Zhang}}, \bibinfo {author} {\bibfnamefont {K.-X.}\ \bibnamefont {Yao}}, \bibinfo {author} {\bibfnamefont {L.}~\bibnamefont {Feng}}, \bibinfo {author} {\bibfnamefont {J.}~\bibnamefont {Hu}},\ and\ \bibinfo {author} {\bibfnamefont {C.}~\bibnamefont {Chin}},\ }\bibfield  {title} {\bibinfo {title} {{Pattern formation in a driven Bose--Einstein condensate}},\ }\href {https://doi.org/10.1038/s41567-020-0839-3} {\bibfield  {journal} {\bibinfo  {journal} {{Nature Physics}}\ }\textbf {\bibinfo {volume} {16}},\ \bibinfo {pages} {652} (\bibinfo {year} {2020})}\BibitemShut {NoStop}%
\bibitem [{\citenamefont {Liu}\ \emph {et~al.}(2023)\citenamefont {Liu}, \citenamefont {Zhang}, \citenamefont {Miao}, \citenamefont {Zhao}, \citenamefont {Wang}, \citenamefont {Chen},\ and\ \citenamefont {Hu}}]{Liu2023}%
  \BibitemOpen
  \bibfield  {author} {\bibinfo {author} {\bibfnamefont {Y.}~\bibnamefont {Liu}}, \bibinfo {author} {\bibfnamefont {Z.}~\bibnamefont {Zhang}}, \bibinfo {author} {\bibfnamefont {S.}~\bibnamefont {Miao}}, \bibinfo {author} {\bibfnamefont {Z.}~\bibnamefont {Zhao}}, \bibinfo {author} {\bibfnamefont {H.}~\bibnamefont {Wang}}, \bibinfo {author} {\bibfnamefont {W.}~\bibnamefont {Chen}},\ and\ \bibinfo {author} {\bibfnamefont {J.}~\bibnamefont {Hu}},\ }\bibfield  {title} {\bibinfo {title} {Calibrating the absorption imaging of cold atoms under high magnetic fields},\ }\href {https://doi.org/10.1103/PhysRevApplied.20.014037} {\bibfield  {journal} {\bibinfo  {journal} {{Phys. Rev. Appl.}}\ }\textbf {\bibinfo {volume} {20}},\ \bibinfo {pages} {014037} (\bibinfo {year} {2023})}\BibitemShut {NoStop}%
\bibitem [{\citenamefont {Hanna}\ \emph {et~al.}(2007)\citenamefont {Hanna}, \citenamefont {K\"ohler},\ and\ \citenamefont {Burnett}}]{Hanna2007}%
  \BibitemOpen
  \bibfield  {author} {\bibinfo {author} {\bibfnamefont {T.~M.}\ \bibnamefont {Hanna}}, \bibinfo {author} {\bibfnamefont {T.}~\bibnamefont {K\"ohler}},\ and\ \bibinfo {author} {\bibfnamefont {K.}~\bibnamefont {Burnett}},\ }\bibfield  {title} {\bibinfo {title} {Association of molecules using a resonantly modulated magnetic field},\ }\href {https://doi.org/10.1103/PhysRevA.75.013606} {\bibfield  {journal} {\bibinfo  {journal} {{Phys. Rev. A}}\ }\textbf {\bibinfo {volume} {75}},\ \bibinfo {pages} {013606} (\bibinfo {year} {2007})}\BibitemShut {NoStop}%
\bibitem [{\citenamefont {Zhao}\ \emph {et~al.}(2021)\citenamefont {Zhao}, \citenamefont {Zhang}, \citenamefont {Chen}, \citenamefont {Wang},\ and\ \citenamefont {Hu}}]{Zhao2021}%
  \BibitemOpen
  \bibfield  {author} {\bibinfo {author} {\bibfnamefont {Y.}~\bibnamefont {Zhao}}, \bibinfo {author} {\bibfnamefont {R.}~\bibnamefont {Zhang}}, \bibinfo {author} {\bibfnamefont {W.}~\bibnamefont {Chen}}, \bibinfo {author} {\bibfnamefont {X.-B.}\ \bibnamefont {Wang}},\ and\ \bibinfo {author} {\bibfnamefont {J.}~\bibnamefont {Hu}},\ }\bibfield  {title} {\bibinfo {title} {{Creation of Greenberger-Horne-Zeilinger states with thousands of atoms by entanglement amplification}},\ }\href {https://doi.org/10.1038/s41534-021-00364-8} {\bibfield  {journal} {\bibinfo  {journal} {{npj Quantum Information}}\ }\textbf {\bibinfo {volume} {7}},\ \bibinfo {pages} {24} (\bibinfo {year} {2021})}\BibitemShut {NoStop}%
\bibitem [{\citenamefont {Venu}\ \emph {et~al.}(2023)\citenamefont {Venu}, \citenamefont {Xu}, \citenamefont {Mamaev}, \citenamefont {Corapi}, \citenamefont {Bilitewski}, \citenamefont {D'Incao}, \citenamefont {Fujiwara}, \citenamefont {Rey},\ and\ \citenamefont {Thywissen}}]{Venu2023}%
  \BibitemOpen
  \bibfield  {author} {\bibinfo {author} {\bibfnamefont {V.}~\bibnamefont {Venu}}, \bibinfo {author} {\bibfnamefont {P.}~\bibnamefont {Xu}}, \bibinfo {author} {\bibfnamefont {M.}~\bibnamefont {Mamaev}}, \bibinfo {author} {\bibfnamefont {F.}~\bibnamefont {Corapi}}, \bibinfo {author} {\bibfnamefont {T.}~\bibnamefont {Bilitewski}}, \bibinfo {author} {\bibfnamefont {J.~P.}\ \bibnamefont {D'Incao}}, \bibinfo {author} {\bibfnamefont {C.~J.}\ \bibnamefont {Fujiwara}}, \bibinfo {author} {\bibfnamefont {A.~M.}\ \bibnamefont {Rey}},\ and\ \bibinfo {author} {\bibfnamefont {J.~H.}\ \bibnamefont {Thywissen}},\ }\bibfield  {title} {\bibinfo {title} {Unitary p-wave interactions between fermions in an optical lattice},\ }\href {https://doi.org/10.1038/s41586-022-05405-6} {\bibfield  {journal} {\bibinfo  {journal} {Nature}\ }\textbf {\bibinfo {volume} {613}},\ \bibinfo {pages} {262} (\bibinfo {year} {2023})}\BibitemShut {NoStop}%
\bibitem [{\citenamefont {Huang}\ \emph {et~al.}(2021)\citenamefont {Huang}, \citenamefont {Yao}, \citenamefont {Liang}, \citenamefont {Wang}, \citenamefont {Zheng}, \citenamefont {Li}, \citenamefont {Xiong}, \citenamefont {Zhou}, \citenamefont {Chen}, \citenamefont {Chen},\ and\ \citenamefont {Hu}}]{Huang2021}%
  \BibitemOpen
  \bibfield  {author} {\bibinfo {author} {\bibfnamefont {Q.}~\bibnamefont {Huang}}, \bibinfo {author} {\bibfnamefont {R.}~\bibnamefont {Yao}}, \bibinfo {author} {\bibfnamefont {L.}~\bibnamefont {Liang}}, \bibinfo {author} {\bibfnamefont {S.}~\bibnamefont {Wang}}, \bibinfo {author} {\bibfnamefont {Q.}~\bibnamefont {Zheng}}, \bibinfo {author} {\bibfnamefont {D.}~\bibnamefont {Li}}, \bibinfo {author} {\bibfnamefont {W.}~\bibnamefont {Xiong}}, \bibinfo {author} {\bibfnamefont {X.}~\bibnamefont {Zhou}}, \bibinfo {author} {\bibfnamefont {W.}~\bibnamefont {Chen}}, \bibinfo {author} {\bibfnamefont {X.}~\bibnamefont {Chen}},\ and\ \bibinfo {author} {\bibfnamefont {J.}~\bibnamefont {Hu}},\ }\bibfield  {title} {\bibinfo {title} {Observation of many-body quantum phase transitions beyond the Kibble-Zurek mechanism},\ }\href {https://doi.org/10.1103/PhysRevLett.127.200601} {\bibfield  {journal} {\bibinfo  {journal} {{Phys. Rev. Lett.}}\ }\textbf {\bibinfo {volume} {127}},\ \bibinfo {pages} {200601} (\bibinfo {year}
  {2021})}\BibitemShut {NoStop}%
\bibitem [{\citenamefont {Yang}\ \emph {et~al.}(2022{\natexlab{b}})\citenamefont {Yang}, \citenamefont {Cao}, \citenamefont {Su}, \citenamefont {Rui}, \citenamefont {Zhao},\ and\ \citenamefont {Pan}}]{Yang2022v2}%
  \BibitemOpen
  \bibfield  {author} {\bibinfo {author} {\bibfnamefont {H.}~\bibnamefont {Yang}}, \bibinfo {author} {\bibfnamefont {J.}~\bibnamefont {Cao}}, \bibinfo {author} {\bibfnamefont {Z.}~\bibnamefont {Su}}, \bibinfo {author} {\bibfnamefont {J.}~\bibnamefont {Rui}}, \bibinfo {author} {\bibfnamefont {B.}~\bibnamefont {Zhao}},\ and\ \bibinfo {author} {\bibfnamefont {J.-W.}\ \bibnamefont {Pan}},\ }\bibfield  {title} {\bibinfo {title} {Creation of an ultracold gas of triatomic molecules from an atom–diatomic molecule mixture},\ }\href {https://doi.org/10.1126/science.ade6307} {\bibfield  {journal} {\bibinfo  {journal} {Science}\ }\textbf {\bibinfo {volume} {378}},\ \bibinfo {pages} {1009} (\bibinfo {year} {2022}{\natexlab{b}})}\BibitemShut {NoStop}%
\bibitem [{\citenamefont {Zhang}\ \emph {et~al.}(2024)\citenamefont {Zhang}, \citenamefont {Chi},\ and\ \citenamefont {Hu}}]{Zhang2024}%
  \BibitemOpen
  \bibfield  {author} {\bibinfo {author} {\bibfnamefont {T.}~\bibnamefont {Zhang}}, \bibinfo {author} {\bibfnamefont {Z.}~\bibnamefont {Chi}},\ and\ \bibinfo {author} {\bibfnamefont {J.}~\bibnamefont {Hu}},\ }\bibfield  {title} {\bibinfo {title} {Entanglement generation via single-qubit rotations in a torn Hilbert space},\ }\href {https://doi.org/10.1103/PRXQuantum.5.030345} {\bibfield  {journal} {\bibinfo  {journal} {{PRX Quantum}}\ }\textbf {\bibinfo {volume} {5}},\ \bibinfo {pages} {030345} (\bibinfo {year} {2024})}\BibitemShut {NoStop}%
\bibitem [{\citenamefont {Theis}\ \emph {et~al.}(2004)\citenamefont {Theis}, \citenamefont {Thalhammer}, \citenamefont {Winkler}, \citenamefont {Hellwig}, \citenamefont {Ruff}, \citenamefont {Grimm},\ and\ \citenamefont {Denschlag}}]{Theis2004}%
  \BibitemOpen
  \bibfield  {author} {\bibinfo {author} {\bibfnamefont {M.}~\bibnamefont {Theis}}, \bibinfo {author} {\bibfnamefont {G.}~\bibnamefont {Thalhammer}}, \bibinfo {author} {\bibfnamefont {K.}~\bibnamefont {Winkler}}, \bibinfo {author} {\bibfnamefont {M.}~\bibnamefont {Hellwig}}, \bibinfo {author} {\bibfnamefont {G.}~\bibnamefont {Ruff}}, \bibinfo {author} {\bibfnamefont {R.}~\bibnamefont {Grimm}},\ and\ \bibinfo {author} {\bibfnamefont {J.~H.}\ \bibnamefont {Denschlag}},\ }\bibfield  {title} {\bibinfo {title} {Tuning the scattering length with an optically induced {F}eshbach resonance},\ }\href {https://doi.org/10.1103/PhysRevLett.93.123001} {\bibfield  {journal} {\bibinfo  {journal} {{Phys. Rev. Lett.}}\ }\textbf {\bibinfo {volume} {93}},\ \bibinfo {pages} {123001} (\bibinfo {year} {2004})}\BibitemShut {NoStop}%
\bibitem [{\citenamefont {Bauer}\ \emph {et~al.}(2009)\citenamefont {Bauer}, \citenamefont {Lettner}, \citenamefont {Vo}, \citenamefont {Rempe},\ and\ \citenamefont {D{\"{u}}rr}}]{Bauer2009}%
  \BibitemOpen
  \bibfield  {author} {\bibinfo {author} {\bibfnamefont {D.~M.}\ \bibnamefont {Bauer}}, \bibinfo {author} {\bibfnamefont {M.}~\bibnamefont {Lettner}}, \bibinfo {author} {\bibfnamefont {C.}~\bibnamefont {Vo}}, \bibinfo {author} {\bibfnamefont {G.}~\bibnamefont {Rempe}},\ and\ \bibinfo {author} {\bibfnamefont {S.}~\bibnamefont {D{\"{u}}rr}},\ }\bibfield  {title} {\bibinfo {title} {{Control of a magnetic Feshbach resonance with laser light}},\ }\href {https://doi.org/10.1038/nphys1232} {\bibfield  {journal} {\bibinfo  {journal} {{Nature Physics}}\ }\textbf {\bibinfo {volume} {5}},\ \bibinfo {pages} {339} (\bibinfo {year} {2009})}\BibitemShut {NoStop}%
\bibitem [{\citenamefont {Clark}\ \emph {et~al.}(2015)\citenamefont {Clark}, \citenamefont {Ha}, \citenamefont {Xu},\ and\ \citenamefont {Chin}}]{PhysRevLett.115.155301}%
  \BibitemOpen
  \bibfield  {author} {\bibinfo {author} {\bibfnamefont {L.~W.}\ \bibnamefont {Clark}}, \bibinfo {author} {\bibfnamefont {L.-C.}\ \bibnamefont {Ha}}, \bibinfo {author} {\bibfnamefont {C.-Y.}\ \bibnamefont {Xu}},\ and\ \bibinfo {author} {\bibfnamefont {C.}~\bibnamefont {Chin}},\ }\bibfield  {title} {\bibinfo {title} {Quantum dynamics with spatiotemporal control of interactions in a stable {B}ose-{E}instein condensate},\ }\href {https://doi.org/10.1103/PhysRevLett.115.155301} {\bibfield  {journal} {\bibinfo  {journal} {{Phys. Rev. Lett.}}\ }\textbf {\bibinfo {volume} {115}},\ \bibinfo {pages} {155301} (\bibinfo {year} {2015})}\BibitemShut {NoStop}%
\bibitem [{\citenamefont {Chen}\ \emph {et~al.}(2023)\citenamefont {Chen}, \citenamefont {Schindewolf}, \citenamefont {Eppelt}, \citenamefont {Bause}, \citenamefont {Duda}, \citenamefont {Biswas}, \citenamefont {Karman}, \citenamefont {Hilker}, \citenamefont {Bloch},\ and\ \citenamefont {Luo}}]{Chen2023}%
  \BibitemOpen
  \bibfield  {author} {\bibinfo {author} {\bibfnamefont {X.-Y.}\ \bibnamefont {Chen}}, \bibinfo {author} {\bibfnamefont {A.}~\bibnamefont {Schindewolf}}, \bibinfo {author} {\bibfnamefont {S.}~\bibnamefont {Eppelt}}, \bibinfo {author} {\bibfnamefont {R.}~\bibnamefont {Bause}}, \bibinfo {author} {\bibfnamefont {M.}~\bibnamefont {Duda}}, \bibinfo {author} {\bibfnamefont {S.}~\bibnamefont {Biswas}}, \bibinfo {author} {\bibfnamefont {T.}~\bibnamefont {Karman}}, \bibinfo {author} {\bibfnamefont {T.}~\bibnamefont {Hilker}}, \bibinfo {author} {\bibfnamefont {I.}~\bibnamefont {Bloch}},\ and\ \bibinfo {author} {\bibfnamefont {X.-Y.}\ \bibnamefont {Luo}},\ }\bibfield  {title} {\bibinfo {title} {Field-linked resonances of polar molecules},\ }\href {https://doi.org/10.1038/s41586-022-05651-8} {\bibfield  {journal} {\bibinfo  {journal} {Nature}\ }\textbf {\bibinfo {volume} {614}},\ \bibinfo {pages} {59} (\bibinfo {year} {2023})}\BibitemShut {NoStop}%
\bibitem [{\citenamefont {Lassabli\`ere}\ and\ \citenamefont {Qu\'em\'ener}(2018)}]{PhysRevLett.121.163402}%
  \BibitemOpen
  \bibfield  {author} {\bibinfo {author} {\bibfnamefont {L.}~\bibnamefont {Lassabli\`ere}}\ and\ \bibinfo {author} {\bibfnamefont {G.}~\bibnamefont {Qu\'em\'ener}},\ }\bibfield  {title} {\bibinfo {title} {Controlling the scattering length of ultracold dipolar molecules},\ }\href {https://doi.org/10.1103/PhysRevLett.121.163402} {\bibfield  {journal} {\bibinfo  {journal} {{Phys. Rev. Lett.}}\ }\textbf {\bibinfo {volume} {121}},\ \bibinfo {pages} {163402} (\bibinfo {year} {2018})}\BibitemShut {NoStop}%
\bibitem [{\citenamefont {Chen}\ \emph {et~al.}(2024)\citenamefont {Chen}, \citenamefont {Biswas}, \citenamefont {Eppelt}, \citenamefont {Schindewolf}, \citenamefont {Deng}, \citenamefont {Shi}, \citenamefont {Yi}, \citenamefont {Hilker}, \citenamefont {Bloch},\ and\ \citenamefont {Luo}}]{Chen2024v2}%
  \BibitemOpen
  \bibfield  {author} {\bibinfo {author} {\bibfnamefont {X.-Y.}\ \bibnamefont {Chen}}, \bibinfo {author} {\bibfnamefont {S.}~\bibnamefont {Biswas}}, \bibinfo {author} {\bibfnamefont {S.}~\bibnamefont {Eppelt}}, \bibinfo {author} {\bibfnamefont {A.}~\bibnamefont {Schindewolf}}, \bibinfo {author} {\bibfnamefont {F.}~\bibnamefont {Deng}}, \bibinfo {author} {\bibfnamefont {T.}~\bibnamefont {Shi}}, \bibinfo {author} {\bibfnamefont {S.}~\bibnamefont {Yi}}, \bibinfo {author} {\bibfnamefont {T.~A.}\ \bibnamefont {Hilker}}, \bibinfo {author} {\bibfnamefont {I.}~\bibnamefont {Bloch}},\ and\ \bibinfo {author} {\bibfnamefont {X.-Y.}\ \bibnamefont {Luo}},\ }\bibfield  {title} {\bibinfo {title} {Ultracold field-linked tetratomic molecules},\ }\href {https://doi.org/10.1038/s41586-023-06986-6} {\bibfield  {journal} {\bibinfo  {journal} {Nature}\ }\textbf {\bibinfo {volume} {626}},\ \bibinfo {pages} {283} (\bibinfo {year} {2024})}\BibitemShut {NoStop}%
\bibitem [{\citenamefont {Nicholson}\ \emph {et~al.}(2015)\citenamefont {Nicholson}, \citenamefont {Blatt}, \citenamefont {Bloom}, \citenamefont {Williams}, \citenamefont {Thomsen}, \citenamefont {Ye},\ and\ \citenamefont {Julienne}}]{PhysRevA.92.022709}%
  \BibitemOpen
  \bibfield  {author} {\bibinfo {author} {\bibfnamefont {T.~L.}\ \bibnamefont {Nicholson}}, \bibinfo {author} {\bibfnamefont {S.}~\bibnamefont {Blatt}}, \bibinfo {author} {\bibfnamefont {B.~J.}\ \bibnamefont {Bloom}}, \bibinfo {author} {\bibfnamefont {J.~R.}\ \bibnamefont {Williams}}, \bibinfo {author} {\bibfnamefont {J.~W.}\ \bibnamefont {Thomsen}}, \bibinfo {author} {\bibfnamefont {J.}~\bibnamefont {Ye}},\ and\ \bibinfo {author} {\bibfnamefont {P.~S.}\ \bibnamefont {Julienne}},\ }\bibfield  {title} {\bibinfo {title} {Optical Feshbach resonances: field-dressed theory and comparison with experiments},\ }\href {https://doi.org/10.1103/PhysRevA.92.022709} {\bibfield  {journal} {\bibinfo  {journal} {Phys. Rev. A}\ }\textbf {\bibinfo {volume} {92}},\ \bibinfo {pages} {022709} (\bibinfo {year} {2015})}\BibitemShut {NoStop}%
\bibitem [{\citenamefont {Fu}\ \emph {et~al.}(2013)\citenamefont {Fu}, \citenamefont {Wang}, \citenamefont {Huang}, \citenamefont {Meng}, \citenamefont {Hu},\ and\ \citenamefont {Zhang}}]{PhysRevA.88.041601}%
  \BibitemOpen
  \bibfield  {author} {\bibinfo {author} {\bibfnamefont {Z.}~\bibnamefont {Fu}}, \bibinfo {author} {\bibfnamefont {P.}~\bibnamefont {Wang}}, \bibinfo {author} {\bibfnamefont {L.}~\bibnamefont {Huang}}, \bibinfo {author} {\bibfnamefont {Z.}~\bibnamefont {Meng}}, \bibinfo {author} {\bibfnamefont {H.}~\bibnamefont {Hu}},\ and\ \bibinfo {author} {\bibfnamefont {J.}~\bibnamefont {Zhang}},\ }\bibfield  {title} {\bibinfo {title} {Optical control of a magnetic Feshbach resonance in an ultracold Fermi gas},\ }\href {https://doi.org/10.1103/PhysRevA.88.041601} {\bibfield  {journal} {\bibinfo  {journal} {Phys. Rev. A}\ }\textbf {\bibinfo {volume} {88}},\ \bibinfo {pages} {041601} (\bibinfo {year} {2013})}\BibitemShut {NoStop}%
\bibitem [{\citenamefont {Papoular}\ \emph {et~al.}(2010)\citenamefont {Papoular}, \citenamefont {Shlyapnikov},\ and\ \citenamefont {Dalibard}}]{PhysRevA.81.041603}%
  \BibitemOpen
  \bibfield  {author} {\bibinfo {author} {\bibfnamefont {D.~J.}\ \bibnamefont {Papoular}}, \bibinfo {author} {\bibfnamefont {G.~V.}\ \bibnamefont {Shlyapnikov}},\ and\ \bibinfo {author} {\bibfnamefont {J.}~\bibnamefont {Dalibard}},\ }\bibfield  {title} {\bibinfo {title} {Microwave-induced Fano-Feshbach resonances},\ }\href {https://doi.org/10.1103/PhysRevA.81.041603} {\bibfield  {journal} {\bibinfo  {journal} {Phys. Rev. A}\ }\textbf {\bibinfo {volume} {81}},\ \bibinfo {pages} {041603} (\bibinfo {year} {2010})}\BibitemShut {NoStop}%
\bibitem [{\citenamefont {Tscherbul}\ \emph {et~al.}(2010)\citenamefont {Tscherbul}, \citenamefont {Calarco}, \citenamefont {Lesanovsky}, \citenamefont {Krems}, \citenamefont {Dalgarno},\ and\ \citenamefont {Schmiedmayer}}]{PhysRevA.81.050701}%
  \BibitemOpen
  \bibfield  {author} {\bibinfo {author} {\bibfnamefont {T.~V.}\ \bibnamefont {Tscherbul}}, \bibinfo {author} {\bibfnamefont {T.}~\bibnamefont {Calarco}}, \bibinfo {author} {\bibfnamefont {I.}~\bibnamefont {Lesanovsky}}, \bibinfo {author} {\bibfnamefont {R.~V.}\ \bibnamefont {Krems}}, \bibinfo {author} {\bibfnamefont {A.}~\bibnamefont {Dalgarno}},\ and\ \bibinfo {author} {\bibfnamefont {J.}~\bibnamefont {Schmiedmayer}},\ }\bibfield  {title} {\bibinfo {title} {Rf-field-induced Feshbach resonances},\ }\href {https://doi.org/10.1103/PhysRevA.81.050701} {\bibfield  {journal} {\bibinfo  {journal} {Phys. Rev. A}\ }\textbf {\bibinfo {volume} {81}},\ \bibinfo {pages} {050701} (\bibinfo {year} {2010})}\BibitemShut {NoStop}%
\bibitem [{\citenamefont {Owens}\ \emph {et~al.}(2016)\citenamefont {Owens}, \citenamefont {Xie},\ and\ \citenamefont {Hutson}}]{PhysRevA.94.023619}%
  \BibitemOpen
  \bibfield  {author} {\bibinfo {author} {\bibfnamefont {D.~J.}\ \bibnamefont {Owens}}, \bibinfo {author} {\bibfnamefont {T.}~\bibnamefont {Xie}},\ and\ \bibinfo {author} {\bibfnamefont {J.~M.}\ \bibnamefont {Hutson}},\ }\bibfield  {title} {\bibinfo {title} {Creating Feshbach resonances for ultracold molecule formation with radio-frequency fields},\ }\href {https://doi.org/10.1103/PhysRevA.94.023619} {\bibfield  {journal} {\bibinfo  {journal} {Phys. Rev. A}\ }\textbf {\bibinfo {volume} {94}},\ \bibinfo {pages} {023619} (\bibinfo {year} {2016})}\BibitemShut {NoStop}%
\bibitem [{\citenamefont {Langmack}\ \emph {et~al.}(2015)\citenamefont {Langmack}, \citenamefont {Smith},\ and\ \citenamefont {Braaten}}]{Langmack2015}%
  \BibitemOpen
  \bibfield  {author} {\bibinfo {author} {\bibfnamefont {C.}~\bibnamefont {Langmack}}, \bibinfo {author} {\bibfnamefont {D.~H.}\ \bibnamefont {Smith}},\ and\ \bibinfo {author} {\bibfnamefont {E.}~\bibnamefont {Braaten}},\ }\bibfield  {title} {\bibinfo {title} {Association of atoms into universal dimers using an oscillating magnetic field},\ }\href {https://doi.org/10.1103/PhysRevLett.114.103002} {\bibfield  {journal} {\bibinfo  {journal} {{Phys. Rev. Lett.}}\ }\textbf {\bibinfo {volume} {114}},\ \bibinfo {pages} {103002} (\bibinfo {year} {2015})}\BibitemShut {NoStop}%
\bibitem [{\citenamefont {Smith}(2015)}]{Hudson2015}%
  \BibitemOpen
  \bibfield  {author} {\bibinfo {author} {\bibfnamefont {D.~H.}\ \bibnamefont {Smith}},\ }\bibfield  {title} {\bibinfo {title} {Inducing resonant interactions in ultracold atoms with a modulated magnetic field},\ }\href {https://doi.org/10.1103/PhysRevLett.115.193002} {\bibfield  {journal} {\bibinfo  {journal} {{Phys. Rev. Lett.}}\ }\textbf {\bibinfo {volume} {115}},\ \bibinfo {pages} {193002} (\bibinfo {year} {2015})}\BibitemShut {NoStop}%
\bibitem [{\citenamefont {Mark}\ \emph {et~al.}(2007)\citenamefont {Mark}, \citenamefont {Ferlaino}, \citenamefont {Knoop}, \citenamefont {Danzl}, \citenamefont {Kraemer}, \citenamefont {Chin}, \citenamefont {N\"agerl},\ and\ \citenamefont {Grimm}}]{Mark2007}%
  \BibitemOpen
  \bibfield  {author} {\bibinfo {author} {\bibfnamefont {M.}~\bibnamefont {Mark}}, \bibinfo {author} {\bibfnamefont {F.}~\bibnamefont {Ferlaino}}, \bibinfo {author} {\bibfnamefont {S.}~\bibnamefont {Knoop}}, \bibinfo {author} {\bibfnamefont {J.~G.}\ \bibnamefont {Danzl}}, \bibinfo {author} {\bibfnamefont {T.}~\bibnamefont {Kraemer}}, \bibinfo {author} {\bibfnamefont {C.}~\bibnamefont {Chin}}, \bibinfo {author} {\bibfnamefont {H.-C.}\ \bibnamefont {N\"agerl}},\ and\ \bibinfo {author} {\bibfnamefont {R.}~\bibnamefont {Grimm}},\ }\bibfield  {title} {\bibinfo {title} {Spectroscopy of ultracold trapped cesium {F}eshbach molecules},\ }\href {https://doi.org/10.1103/PhysRevA.76.042514} {\bibfield  {journal} {\bibinfo  {journal} {{Phys. Rev. A}}\ }\textbf {\bibinfo {volume} {76}},\ \bibinfo {pages} {042514} (\bibinfo {year} {2007})}\BibitemShut {NoStop}%
\bibitem[{\citenamefont {Hsu} \emph{et~al.}(2016)}]{Hsu2016}%
\BibitemOpen
  \bibfield{author} {\bibinfo{author} {\bibfnamefont {C.-W.}\ \bibnamefont {Hsu}, \bibfnamefont {B.}\ \bibnamefont {Zhen}, \bibfnamefont {A.~D.}\ \bibnamefont {Stone}, \bibfnamefont {J.~D.}\ \bibnamefont {Joannopoulos}, \bibfnamefont {and~M.}\ \bibnamefont {Soljačić}}},\ 
  \bibfield{title} {\bibinfo{title} {Bound states in the continuum}},\ 
  \href{https://doi.org/10.1038/natrevmats.2016.48} {%
    \bibfield{journal} {\bibinfo{journal} {Nat.~Rev.~Mater.}}\ 
    \textbf{\bibinfo{volume}{1}},\ 
    \bibinfo{pages}{16048} 
    (\bibinfo{year}{2016})%
  }\BibitemShut{NoStop}%
\bibitem [{\citenamefont {Friedrich}\ and\ \citenamefont {Wintgen}(1985)}]{PhysRevA.32.3231}%
  \BibitemOpen
  \bibfield  {author} {\bibinfo {author} {\bibfnamefont {H.}~\bibnamefont {Friedrich}}\ and\ \bibinfo {author} {\bibfnamefont {D.}~\bibnamefont {Wintgen}},\ }\bibfield  {title} {\bibinfo {title} {Interfering resonances and bound states in the continuum},\ }\href {https://doi.org/10.1103/PhysRevA.32.3231} {\bibfield  {journal} {\bibinfo  {journal} {Phys. Rev. A}\ }\textbf {\bibinfo {volume} {32}},\ \bibinfo {pages} {3231} (\bibinfo {year} {1985})}\BibitemShut {NoStop}%
\bibitem [{\citenamefont {Jagannathan}\ \emph {et~al.}(2016)\citenamefont {Jagannathan}, \citenamefont {Arunkumar}, \citenamefont {Joseph},\ and\ \citenamefont {Thomas}}]{Jagannathan2016}%
  \BibitemOpen
  \bibfield  {author} {\bibinfo {author} {\bibfnamefont {A.}~\bibnamefont {Jagannathan}}, \bibinfo {author} {\bibfnamefont {N.}~\bibnamefont {Arunkumar}}, \bibinfo {author} {\bibfnamefont {J.~A.}\ \bibnamefont {Joseph}},\ and\ \bibinfo {author} {\bibfnamefont {J.~E.}\ \bibnamefont {Thomas}},\ }\bibfield  {title} {\bibinfo {title} {Optical control of magnetic {F}eshbach resonances by closed-channel electromagnetically induced transparency},\ }\href {https://doi.org/10.1103/PhysRevLett.116.075301} {\bibfield  {journal} {\bibinfo  {journal} {{Phys. Rev. Lett.}}\ }\textbf {\bibinfo {volume} {116}},\ \bibinfo {pages} {075301} (\bibinfo {year} {2016})}\BibitemShut {NoStop}%
\bibitem [{\citenamefont {Tanji-Suzuki}\ \emph {et~al.}(2011)\citenamefont {Tanji-Suzuki}, \citenamefont {Chen}, \citenamefont {Landig}, \citenamefont {Simon},\ and\ \citenamefont {Vuletić}}]{Tanji2011}%
  \BibitemOpen
  \bibfield  {author} {\bibinfo {author} {\bibfnamefont {H.}~\bibnamefont {Tanji-Suzuki}}, \bibinfo {author} {\bibfnamefont {W.}~\bibnamefont {Chen}}, \bibinfo {author} {\bibfnamefont {R.}~\bibnamefont {Landig}}, \bibinfo {author} {\bibfnamefont {J.}~\bibnamefont {Simon}},\ and\ \bibinfo {author} {\bibfnamefont {V.}~\bibnamefont {Vuletić}},\ }\bibfield  {title} {\bibinfo {title} {Vacuum-induced transparency},\ }\href {https://doi.org/10.1126/science.1208066} {\bibfield  {journal} {\bibinfo  {journal} {Science}\ }\textbf {\bibinfo {volume} {333}},\ \bibinfo {pages} {1266} (\bibinfo {year} {2011})}\BibitemShut {NoStop}%
\bibitem [{\citenamefont {Chen}\ \emph {et~al.}(2013)\citenamefont {Chen}, \citenamefont {Beck}, \citenamefont {Bücker}, \citenamefont {Gullans}, \citenamefont {Lukin}, \citenamefont {Tanji-Suzuki},\ and\ \citenamefont {Vuletić}}]{Wenlan2013}%
  \BibitemOpen
  \bibfield  {author} {\bibinfo {author} {\bibfnamefont {W.}~\bibnamefont {Chen}}, \bibinfo {author} {\bibfnamefont {K.~M.}\ \bibnamefont {Beck}}, \bibinfo {author} {\bibfnamefont {R.}~\bibnamefont {Bücker}}, \bibinfo {author} {\bibfnamefont {M.}~\bibnamefont {Gullans}}, \bibinfo {author} {\bibfnamefont {M.~D.}\ \bibnamefont {Lukin}}, \bibinfo {author} {\bibfnamefont {H.}~\bibnamefont {Tanji-Suzuki}},\ and\ \bibinfo {author} {\bibfnamefont {V.}~\bibnamefont {Vuletić}},\ }\bibfield  {title} {\bibinfo {title} {All-optical switch and transistor gated by one stored photon},\ }\href {https://doi.org/10.1126/science.1238169} {\bibfield  {journal} {\bibinfo  {journal} {Science}\ }\textbf {\bibinfo {volume} {341}},\ \bibinfo {pages} {768} (\bibinfo {year} {2013})}\BibitemShut {NoStop}%
\bibitem [{\citenamefont {Sun}\ \emph {et~al.}(2018)\citenamefont {Sun}, \citenamefont {Kim}, \citenamefont {Luo}, \citenamefont {Solomon},\ and\ \citenamefont {Waks}}]{Sun2018}%
  \BibitemOpen
  \bibfield  {author} {\bibinfo {author} {\bibfnamefont {S.}~\bibnamefont {Sun}}, \bibinfo {author} {\bibfnamefont {H.}~\bibnamefont {Kim}}, \bibinfo {author} {\bibfnamefont {Z.}~\bibnamefont {Luo}}, \bibinfo {author} {\bibfnamefont {G.~S.}\ \bibnamefont {Solomon}},\ and\ \bibinfo {author} {\bibfnamefont {E.}~\bibnamefont {Waks}},\ }\bibfield  {title} {\bibinfo {title} {A single-photon switch and transistor enabled by a solid-state quantum memory},\ }\href {https://doi.org/10.1126/science.aat3581} {\bibfield  {journal} {\bibinfo  {journal} {Science}\ }\textbf {\bibinfo {volume} {361}},\ \bibinfo {pages} {57} (\bibinfo {year} {2018})}\BibitemShut {NoStop}%
\bibitem [{\citenamefont {Niemietz}\ \emph {et~al.}(2021)\citenamefont {Niemietz}, \citenamefont {Farrera}, \citenamefont {Langenfeld},\ and\ \citenamefont {Rempe}}]{Niemietz2021}%
  \BibitemOpen
  \bibfield  {author} {\bibinfo {author} {\bibfnamefont {D.}~\bibnamefont {Niemietz}}, \bibinfo {author} {\bibfnamefont {P.}~\bibnamefont {Farrera}}, \bibinfo {author} {\bibfnamefont {S.}~\bibnamefont {Langenfeld}},\ and\ \bibinfo {author} {\bibfnamefont {G.}~\bibnamefont {Rempe}},\ }\bibfield  {title} {\bibinfo {title} {Nondestructive detection of photonic qubits},\ }\href {https://doi.org/10.1038/s41586-021-03290-z} {\bibfield  {journal} {\bibinfo  {journal} {Nature}\ }\textbf {\bibinfo {volume} {591}},\ \bibinfo {pages} {570} (\bibinfo {year} {2021})}\BibitemShut {NoStop}%
\bibitem [{\citenamefont {Kuhn}\ \emph {et~al.}(2002)\citenamefont {Kuhn}, \citenamefont {Hennrich},\ and\ \citenamefont {Rempe}}]{PhysRevLett.89.067901}%
  \BibitemOpen
  \bibfield  {author} {\bibinfo {author} {\bibfnamefont {A.}~\bibnamefont {Kuhn}}, \bibinfo {author} {\bibfnamefont {M.}~\bibnamefont {Hennrich}},\ and\ \bibinfo {author} {\bibfnamefont {G.}~\bibnamefont {Rempe}},\ }\bibfield  {title} {\bibinfo {title} {Deterministic single-photon source for distributed quantum networking},\ }\href {https://doi.org/10.1103/PhysRevLett.89.067901} {\bibfield  {journal} {\bibinfo  {journal} {Phys. Rev. Lett.}\ }\textbf {\bibinfo {volume} {89}},\ \bibinfo {pages} {067901} (\bibinfo {year} {2002})}\BibitemShut {NoStop}%
\bibitem [{\citenamefont {Wang}\ \emph {et~al.}(2019)\citenamefont {Wang}, \citenamefont {Li}, \citenamefont {Zhang}, \citenamefont {Su}, \citenamefont {Zhou}, \citenamefont {Liao}, \citenamefont {Du}, \citenamefont {Yan},\ and\ \citenamefont {Zhu}}]{Wang20199}%
  \BibitemOpen
  \bibfield  {author} {\bibinfo {author} {\bibfnamefont {Y.}~\bibnamefont {Wang}}, \bibinfo {author} {\bibfnamefont {J.}~\bibnamefont {Li}}, \bibinfo {author} {\bibfnamefont {S.}~\bibnamefont {Zhang}}, \bibinfo {author} {\bibfnamefont {K.}~\bibnamefont {Su}}, \bibinfo {author} {\bibfnamefont {Y.}~\bibnamefont {Zhou}}, \bibinfo {author} {\bibfnamefont {K.}~\bibnamefont {Liao}}, \bibinfo {author} {\bibfnamefont {S.}~\bibnamefont {Du}}, \bibinfo {author} {\bibfnamefont {H.}~\bibnamefont {Yan}},\ and\ \bibinfo {author} {\bibfnamefont {S.-L.}\ \bibnamefont {Zhu}},\ }\bibfield  {title} {\bibinfo {title} {Efficient quantum memory for single-photon polarization qubits},\ }\href {https://doi.org/10.1038/s41566-019-0368-8} {\bibfield  {journal} {\bibinfo  {journal} {Nature Photonics}\ }\textbf {\bibinfo {volume} {13}},\ \bibinfo {pages} {346} (\bibinfo {year} {2019})}\BibitemShut {NoStop}%
\bibitem [{Sup(2025)}]{Supplementary}%
  \BibitemOpen
  \bibfield  {title} {{\bibinfo {title} {See Supplemental Material at [url] for additional experimental details, as well as theoretical analysis of the scattering properties under modulation, which includes Ref. [46-57]}}\
  }\BibitemShut {NoStop}%
\bibitem [{\citenamefont {Le~Kien}\ \emph {et~al.}(2013)\citenamefont {Le~Kien}, \citenamefont {Schneeweiss},\ and\ \citenamefont {Rauschenbeutel}}]{2013Dynamical}%
  \BibitemOpen
  \bibfield  {author} {\bibinfo {author} {\bibfnamefont {F.}~\bibnamefont {Le~Kien}}, \bibinfo {author} {\bibfnamefont {P.}~\bibnamefont {Schneeweiss}},\ and\ \bibinfo {author} {\bibfnamefont {A.}~\bibnamefont {Rauschenbeutel}},\ }\bibfield  {title} {\bibinfo {title} {Dynamical polarizability of atoms in arbitrary light fields: general theory and application to cesium},\ }\href {https://doi.org/10.1140/epjd/e2013-30729-x} {\bibfield  {journal} {\bibinfo  {journal} {Eur. Phys. J. D}\ }\textbf {\bibinfo {volume} {67}},\ \bibinfo {pages} {92} (\bibinfo {year} {2013})}\BibitemShut {NoStop}%
\bibitem [{\citenamefont {Wang}\ \emph {et~al.}(2024)\citenamefont {Wang}, \citenamefont {Dong}, \citenamefont {Wang}, \citenamefont {Zhou}, \citenamefont {Huang}, \citenamefont {Zuo}, \citenamefont {Siviloglou},\ and\ \citenamefont {Chen}}]{PhysRevResearch.6.L042002}%
  \BibitemOpen
  \bibfield  {author} {\bibinfo {author} {\bibfnamefont {J.}~\bibnamefont {Wang}}, \bibinfo {author} {\bibfnamefont {L.}~\bibnamefont {Dong}}, \bibinfo {author} {\bibfnamefont {X.}~\bibnamefont {Wang}}, \bibinfo {author} {\bibfnamefont {Z.}~\bibnamefont {Zhou}}, \bibinfo {author} {\bibfnamefont {J.}~\bibnamefont {Huang}}, \bibinfo {author} {\bibfnamefont {Y.}~\bibnamefont {Zuo}}, \bibinfo {author} {\bibfnamefont {G.~A.}\ \bibnamefont {Siviloglou}},\ and\ \bibinfo {author} {\bibfnamefont {J.~F.}\ \bibnamefont {Chen}},\ }\bibfield  {title} {\bibinfo {title} {Light-induced fictitious magnetic fields for quantum storage in cold atomic ensembles},\ }\href {https://doi.org/10.1103/PhysRevResearch.6.L042002} {\bibfield  {journal} {\bibinfo  {journal} {Phys. Rev. Res.}\ }\textbf {\bibinfo {volume} {6}},\ \bibinfo {pages} {L042002} (\bibinfo {year} {2024})}\BibitemShut {NoStop}%
\bibitem [{\citenamefont {K\"ohler}\ \emph {et~al.}(2006)\citenamefont {K\"ohler}, \citenamefont {G\'oral},\ and\ \citenamefont {Julienne}}]{RevModPhys.78.1311}%
  \BibitemOpen
  \bibfield  {author} {\bibinfo {author} {\bibfnamefont {T.}~\bibnamefont {K\"ohler}}, \bibinfo {author} {\bibfnamefont {K.}~\bibnamefont {G\'oral}},\ and\ \bibinfo {author} {\bibfnamefont {P.~S.}\ \bibnamefont {Julienne}},\ }\bibfield  {title} {\bibinfo {title} {Production of cold molecules via magnetically tunable Feshbach resonances},\ }\href {https://doi.org/10.1103/RevModPhys.78.1311} {\bibfield  {journal} {\bibinfo  {journal} {Rev. Mod. Phys.}\ }\textbf {\bibinfo {volume} {78}},\ \bibinfo {pages} {1311} (\bibinfo {year} {2006})}\BibitemShut {NoStop}%
\bibitem [{\citenamefont {Holthaus}(2015)}]{Holthaus_2016}%
  \BibitemOpen
  \bibfield  {author} {\bibinfo {author} {\bibfnamefont {M.}~\bibnamefont {Holthaus}},\ }\bibfield  {title} {\bibinfo {title} {Floquet engineering with quasienergy bands of periodically driven optical lattices},\ }\href {https://doi.org/10.1088/0953-4075/49/1/013001} {\bibfield  {journal} {\bibinfo  {journal} {J. Phys. B: At. Mol. Opt. Phys.}\ }\textbf {\bibinfo {volume} {49}},\ \bibinfo {pages} {013001} (\bibinfo {year} {2015})}\BibitemShut {NoStop}%
\bibitem [{\citenamefont {Kaufman}\ \emph {et~al.}(2009)\citenamefont {Kaufman}, \citenamefont {Anderson}, \citenamefont {Hanna}, \citenamefont {Tiesinga}, \citenamefont {Julienne},\ and\ \citenamefont {Hall}}]{PhysRevA.80.050701}%
  \BibitemOpen
  \bibfield  {author} {\bibinfo {author} {\bibfnamefont {A.~M.}\ \bibnamefont {Kaufman}}, \bibinfo {author} {\bibfnamefont {R.~P.}\ \bibnamefont {Anderson}}, \bibinfo {author} {\bibfnamefont {T.~M.}\ \bibnamefont {Hanna}}, \bibinfo {author} {\bibfnamefont {E.}~\bibnamefont {Tiesinga}}, \bibinfo {author} {\bibfnamefont {P.~S.}\ \bibnamefont {Julienne}},\ and\ \bibinfo {author} {\bibfnamefont {D.~S.}\ \bibnamefont {Hall}},\ }\bibfield  {title} {\bibinfo {title} {Radio-frequency dressing of multiple Feshbach resonances},\ }\href {https://doi.org/10.1103/PhysRevA.80.050701} {\bibfield  {journal} {\bibinfo  {journal} {Phys. Rev. A}\ }\textbf {\bibinfo {volume} {80}},\ \bibinfo {pages} {050701} (\bibinfo {year} {2009})}\BibitemShut {NoStop}%
\bibitem [{\citenamefont {Cohen-Tannoudji}\ \emph {et~al.}()\citenamefont {Cohen-Tannoudji}, \citenamefont {Dupont-Roc},\ and\ \citenamefont {Grynberg}}]{cohen1998atom}%
  \BibitemOpen
  \bibfield  {author} {\bibinfo {author} {\bibfnamefont {C.}~\bibnamefont {Cohen-Tannoudji}}, \bibinfo {author} {\bibfnamefont {J.}~\bibnamefont {Dupont-Roc}},\ and\ \bibinfo {author} {\bibfnamefont {G.}~\bibnamefont {Grynberg}},\ }\href@noop {} {\bibinfo {title} {\textit{Atom-Photon Interactions: Basic Processes and Applications} ({W}iley, {N}ew {Y}ork, 1992).}}\BibitemShut {Stop}%
\bibitem [{\citenamefont {Bohn}\ and\ \citenamefont {Julienne}(1999)}]{PhysRevA.60.414}%
  \BibitemOpen
  \bibfield  {author} {\bibinfo {author} {\bibfnamefont {J.~L.}\ \bibnamefont {Bohn}}\ and\ \bibinfo {author} {\bibfnamefont {P.~S.}\ \bibnamefont {Julienne}},\ }\bibfield  {title} {\bibinfo {title} {Semianalytic theory of laser-assisted resonant cold collisions},\ }\href {https://doi.org/10.1103/PhysRevA.60.414} {\bibfield  {journal} {\bibinfo  {journal} {Phys. Rev. A}\ }\textbf {\bibinfo {volume} {60}},\ \bibinfo {pages} {414} (\bibinfo {year} {1999})}\BibitemShut {NoStop}%
\bibitem [{\citenamefont {Shirley}(1965)}]{PhysRev.138.B979}%
  \BibitemOpen
  \bibfield  {author} {\bibinfo {author} {\bibfnamefont {J.~H.}\ \bibnamefont {Shirley}},\ }\bibfield  {title} {\bibinfo {title} {Solution of the Schr\"odinger equation with a Hamiltonian periodic in time},\ }\href {https://doi.org/10.1103/PhysRev.138.B979} {\bibfield  {journal} {\bibinfo  {journal} {Phys. Rev.}\ }\textbf {\bibinfo {volume} {138}},\ \bibinfo {pages} {B979} (\bibinfo {year} {1965})}\BibitemShut {NoStop}%
\bibitem [{\citenamefont {Viebahn}\ \emph {et~al.}(2021)\citenamefont {Viebahn}, \citenamefont {Minguzzi}, \citenamefont {Sandholzer}, \citenamefont {Walter}, \citenamefont {Sajnani}, \citenamefont {G\"org},\ and\ \citenamefont {Esslinger}}]{PhysRevX.11.011057}%
  \BibitemOpen
  \bibfield  {author} {\bibinfo {author} {\bibfnamefont {K.}~\bibnamefont {Viebahn}}, \bibinfo {author} {\bibfnamefont {J.}~\bibnamefont {Minguzzi}}, \bibinfo {author} {\bibfnamefont {K.}~\bibnamefont {Sandholzer}}, \bibinfo {author} {\bibfnamefont {A.-S.}\ \bibnamefont {Walter}}, \bibinfo {author} {\bibfnamefont {M.}~\bibnamefont {Sajnani}}, \bibinfo {author} {\bibfnamefont {F.}~\bibnamefont {G\"org}},\ and\ \bibinfo {author} {\bibfnamefont {T.}~\bibnamefont {Esslinger}},\ }\bibfield  {title} {\bibinfo {title} {Suppressing dissipation in a Floquet-Hubbard system},\ }\href {https://doi.org/10.1103/PhysRevX.11.011057} {\bibfield  {journal} {\bibinfo  {journal} {Phys. Rev. X}\ }\textbf {\bibinfo {volume} {11}},\ \bibinfo {pages} {011057} (\bibinfo {year} {2021})}\BibitemShut {NoStop}%
\bibitem [{\citenamefont {Chen}\ \emph {et~al.}(2025)\citenamefont {Chen}, \citenamefont {Zhu},\ and\ \citenamefont {Viebahn}}]{tzg7-v7nh}%
  \BibitemOpen
  \bibfield  {author} {\bibinfo {author} {\bibfnamefont {Y.}~\bibnamefont {Chen}}, \bibinfo {author} {\bibfnamefont {Z.}~\bibnamefont {Zhu}},\ and\ \bibinfo {author} {\bibfnamefont {K.}~\bibnamefont {Viebahn}},\ }\bibfield  {title} {\bibinfo {title} {Mitigating higher-band heating in Floquet-Hubbard lattices via two-tone driving},\ }\href {https://doi.org/10.1103/tzg7-v7nh} {\bibfield  {journal} {\bibinfo  {journal} {Phys. Rev. A}\ }\textbf {\bibinfo {volume} {112}},\ \bibinfo {pages} {L021301} (\bibinfo {year} {2025})}\BibitemShut {NoStop}%
\bibitem [{\citenamefont {Guthmann}\ \emph {et~al.}(2025)\citenamefont {Guthmann}, \citenamefont {Lang}, \citenamefont {Kienesberger}, \citenamefont {Barbosa},\ and\ \citenamefont {Widera}}]{doi:10.1126/sciadv.adw3856}%
  \BibitemOpen
  \bibfield  {author} {\bibinfo {author} {\bibfnamefont {A.}~\bibnamefont {Guthmann}}, \bibinfo {author} {\bibfnamefont {F.}~\bibnamefont {Lang}}, \bibinfo {author} {\bibfnamefont {L.~M.}\ \bibnamefont {Kienesberger}}, \bibinfo {author} {\bibfnamefont {S.}~\bibnamefont {Barbosa}},\ and\ \bibinfo {author} {\bibfnamefont {A.}~\bibnamefont {Widera}},\ }\bibfield  {title} {\bibinfo {title} {Floquet engineering of Feshbach resonances in ultracold gases},\ }\href {https://doi.org/10.1126/sciadv.adw3856} {\bibfield  {journal} {\bibinfo  {journal} {Science Advances}\ }\textbf {\bibinfo {volume} {11}},\ \bibinfo {pages} {eadw3856} (\bibinfo {year} {2025})}\BibitemShut {NoStop}%
\bibitem [{\citenamefont {Steck}()}]{Steck}%
  \BibitemOpen
  \bibfield  {author} {\bibinfo {author} {\bibfnamefont {D.~A.}\ \bibnamefont {Steck}},\ }\href@noop {} {\bibinfo {title} {Quantum and atom optics}},\ \bibinfo {howpublished} {available online at \url{http://steck.us/teaching} (revision 0.16.2, 15 November 2024)}\BibitemShut {NoStop}%
\bibitem [{\citenamefont {Berninger}\ \emph {et~al.}(2013)\citenamefont {Berninger}, \citenamefont {Zenesini}, \citenamefont {Huang}, \citenamefont {Harm}, \citenamefont {N\"agerl}, \citenamefont {Ferlaino}, \citenamefont {Grimm}, \citenamefont {Julienne},\ and\ \citenamefont {Hutson}}]{2012model}%
  \BibitemOpen
  \bibfield  {author} {\bibinfo {author} {\bibfnamefont {M.}~\bibnamefont {Berninger}}, \bibinfo {author} {\bibfnamefont {A.}~\bibnamefont {Zenesini}}, \bibinfo {author} {\bibfnamefont {B.}~\bibnamefont {Huang}}, \bibinfo {author} {\bibfnamefont {W.}~\bibnamefont {Harm}}, \bibinfo {author} {\bibfnamefont {H.-C.}\ \bibnamefont {N\"agerl}}, \bibinfo {author} {\bibfnamefont {F.}~\bibnamefont {Ferlaino}}, \bibinfo {author} {\bibfnamefont {R.}~\bibnamefont {Grimm}}, \bibinfo {author} {\bibfnamefont {P.~S.}\ \bibnamefont {Julienne}},\ and\ \bibinfo {author} {\bibfnamefont {J.~M.}\ \bibnamefont {Hutson}},\ }\bibfield  {title} {\bibinfo {title} {Feshbach resonances, weakly bound molecular states, and coupled-channel potentials for cesium at high magnetic fields},\ }\href {https://doi.org/10.1103/PhysRevA.87.032517} {\bibfield  {journal} {\bibinfo  {journal} {Phys. Rev. A}\ }\textbf {\bibinfo {volume} {87}},\ \bibinfo {pages} {032517} (\bibinfo {year} {2013})}\BibitemShut {NoStop}%
\bibitem [{\citenamefont {Lange}\ \emph {et~al.}(2009)\citenamefont {Lange}, \citenamefont {Pilch}, \citenamefont {Prantner}, \citenamefont {Ferlaino}, \citenamefont {Engeser}, \citenamefont {N\"agerl}, \citenamefont {Grimm},\ and\ \citenamefont {Chin}}]{2009fitting}%
  \BibitemOpen
  \bibfield  {author} {\bibinfo {author} {\bibfnamefont {A.~D.}\ \bibnamefont {Lange}}, \bibinfo {author} {\bibfnamefont {K.}~\bibnamefont {Pilch}}, \bibinfo {author} {\bibfnamefont {A.}~\bibnamefont {Prantner}}, \bibinfo {author} {\bibfnamefont {F.}~\bibnamefont {Ferlaino}}, \bibinfo {author} {\bibfnamefont {B.}~\bibnamefont {Engeser}}, \bibinfo {author} {\bibfnamefont {H.-C.}\ \bibnamefont {N\"agerl}}, \bibinfo {author} {\bibfnamefont {R.}~\bibnamefont {Grimm}},\ and\ \bibinfo {author} {\bibfnamefont {C.}~\bibnamefont {Chin}},\ }\bibfield  {title} {\bibinfo {title} {Determination of atomic scattering lengths from measurements of molecular binding energies near Feshbach resonances},\ }\href {https://doi.org/10.1103/PhysRevA.79.013622} {\bibfield  {journal} {\bibinfo  {journal} {Phys. Rev. A}\ }\textbf {\bibinfo {volume} {79}},\ \bibinfo {pages} {013622} (\bibinfo {year} {2009})}\BibitemShut {NoStop}%
\bibitem [{\citenamefont {Ospelkaus}\ \emph {et~al.}(2006)\citenamefont {Ospelkaus}, \citenamefont {Ospelkaus}, \citenamefont {Humbert}, \citenamefont {Ernst}, \citenamefont {Sengstock},\ and\ \citenamefont {Bongs}}]{PhysRevLett.97.120402}%
  \BibitemOpen
  \bibfield  {author} {\bibinfo {author} {\bibfnamefont {C.}~\bibnamefont {Ospelkaus}}, \bibinfo {author} {\bibfnamefont {S.}~\bibnamefont {Ospelkaus}}, \bibinfo {author} {\bibfnamefont {L.}~\bibnamefont {Humbert}}, \bibinfo {author} {\bibfnamefont {P.}~\bibnamefont {Ernst}}, \bibinfo {author} {\bibfnamefont {K.}~\bibnamefont {Sengstock}},\ and\ \bibinfo {author} {\bibfnamefont {K.}~\bibnamefont {Bongs}},\ }\bibfield  {title} {\bibinfo {title} {Ultracold heteronuclear molecules in a 3D optical lattice},\ }\href {https://doi.org/10.1103/PhysRevLett.97.120402} {\bibfield  {journal} {\bibinfo  {journal} {Phys. Rev. Lett.}\ }\textbf {\bibinfo {volume} {97}},\ \bibinfo {pages} {120402} (\bibinfo {year} {2006})}\BibitemShut {NoStop}%
\bibitem [{\citenamefont {Thompson}\ \emph {et~al.}(2005)\citenamefont {Thompson}, \citenamefont {Hodby},\ and\ \citenamefont {Wieman}}]{PhysRevLett.95.190404}%
  \BibitemOpen
  \bibfield  {author} {\bibinfo {author} {\bibfnamefont {S.~T.}\ \bibnamefont {Thompson}}, \bibinfo {author} {\bibfnamefont {E.}~\bibnamefont {Hodby}},\ and\ \bibinfo {author} {\bibfnamefont {C.~E.}\ \bibnamefont {Wieman}},\ }\bibfield  {title} {\bibinfo {title} {Ultracold molecule production via a resonant oscillating magnetic field},\ }\href {https://doi.org/10.1103/PhysRevLett.95.190404} {\bibfield  {journal} {\bibinfo  {journal} {Phys. Rev. Lett.}\ }\textbf {\bibinfo {volume} {95}},\ \bibinfo {pages} {190404} (\bibinfo {year} {2005})}\BibitemShut {NoStop}%
\end{thebibliography}
\end{document}

%\title{Title}
%\author{You name here}
%\affiliation{Somewhere}
%\date{\today}
%\pacs{}
%\maketitle

%%%%%%%%%% Merge with supplemental materials %%%%%%%%%%

\begin{center}
\textbf{\large Supplementary Material for: Observation of modulation-induced Feshbach resonance}
\end{center}
%%%%%%%%%% Merge with supplemental materials %%%%%%%%%%
%%%%%%%%%% Prefix a "S" to all equations, figures, tables and reset the counter %%%%%%%%%%
\setcounter{equation}{0}
\setcounter{figure}{0}
\setcounter{table}{0}
\setcounter{page}{1}
\makeatletter
\renewcommand{\theequation}{S\arabic{equation}}
\renewcommand{\thefigure}{S\arabic{figure}}
\renewcommand{\bibnumfmt}[1]{[S#1]}
\renewcommand{\citenumfont}[1]{S#1}
%%%%%%%%%% Prefix a "S" to all equations, figures, tables and reset the counter %%%%%%%%%%
\section{EXPERIMENTAL SETUP}
Our experiments begin by preparing an almost pure Bose-Einstein condensate (BEC) of approximately $10^5$ cesium atoms in the hyperfine state $\ket{3,3}$. These atoms are confined in an optical trap consisting of a light sheet and a cross-dipole trap formed by a 1064-nm laser in the $x$-$y$ plane. Under a uniform magnetic field, the typical trap frequencies are $\omega_z = 2\pi \times 65\,\text{Hz}$ (vertical) and $\omega_r = 2\pi \times 28\,\text{Hz}$ (horizontal). The final evaporation field is set to $20.8\,\text{G}$, where the scattering length is small and positive ($a = 200\,a_0$).
We then ramp the magnetic field to a target value (stable to within $4\,\text{mG}$, calibrated via microwave) and switch on the modulation light. After an optimized hold time ($1$--$20\,\text{ms}$), we simultaneously extinguish the light and ramp the field back to $20.8\,\text{G}$ for expansion and imaging. Finally, absorption imaging measures the remaining atom number after $18\,\text{ms}$ of time-of-flight.

Light intensity modulation is achieved through the interference of two laser beams, each passing through a separate acousto-optic modulator(AOM) operating at different frequencies, combined by a beam splitter. The variation of light intensity over time is measured by a biased Silicon detector with the band width about 80MHz, which shows that the modulation depth is about 86$\%$, while the half-width of modulation frequency is lower than 1 Hz. The beam waist of modulation light is chosen to be 844 $\mu m$, which is much larger than the typical Bose-Einstein condensate size, and the average peak intensity is ranged from 0.2 W/cm$^2$ to 1.7  W/cm$^2$ according to experimental requirements.

When scanning the energy spectrum in the main text, the binding energy detection for the $4g4$, $6s$, and $6g(6)$ states starts from an almost pure Bose-Einstein condensate as described above. However, since the scattering length is $\sim 900\,a_0$ in the region of $47$--$48.5\,\text{G}$, the $4d$ data are measured by initializing a mixture of Bose-Einstein condensate and thermal atoms to reduce three-body recombination losses, which improves the signal-to-noise ratio of atom loss peaks. 

\textcolor{black}{The resolution of our measured resonance frequencies is estimated to be $\sim$ 10 kHz. Several effects contribute to this resolution. First of all, the measured frequencies are all shifted due to the applied light field. This shift is composed of a dominant linear DC component of light shift, with a negligible contribution from modulation effect. We compensate for this shift through linear fitting between the atom loss peak position and the corresponding average peak intensity. This correction introduces an uncertainty primarily limited by the total power fluctuation, resulting in an estimated error of 2 kHz. Besides, the mean-field shift remains below 1 kHz for our estimated parameters of $\sim 10^{13}$ cm$^{-3}$ density and $1000a_0$ scattering length. The thermal shift is evaluated to be within $k_B\times100$ nK, corresponding to $\sim$ 2 kHz. In the end, the $\sim$ 4 mG magnetic field instability results in the accuracy of $\sim$ 10 kHz, estimated with a maximum magnetic moment of 2.3 MHz/G in experiments.}

\section{Light shift of free-scattering states and molecular states}
In our experiments, the 23GHz red-detuned, $\sigma^-$-polarized light generates different light shifts on the free scattering state of two atoms in the $\ket{F = 3, m_F = 3}$ state and the molecular states, as described in the main text. Especially, the Feshbach resonance at 19.84 G is shifted to 19.70 G under a 0.87 W/cm$^2$ peak intensity light (the shift is same for the light without modulation). The observed 140 mG shift is beyond the region where the shift is mainly attributed to the fictitious magnetic field as reported in \cite{Ref1} with a much larger detuning light.

To make a comparison, we calculate the fictitious magnetic field \cite{Ref2,Ref5} :
\begin{equation}
    \mathbf{B}_z^{f} = -\frac{I \mu_0 c}{2 \mu_B g_F F} ({|\mathbf{u}_{-1}|}^2-{|\mathbf{u}_{+1}|^2})\alpha_v, \label{SupEq1}
\end{equation}
where $I$ is the intensity of light, $\mu_0$ is the vacuum permeability, $c$ is the speed of light, $\mu_B$ is the Bhor magneton, $g_F$ is the hyperfine Land\'{e} g-factor, $\mathbf{u} = \sum_{q = 0,\pm 1} (-1)^q \mathbf{u}_{q} \hat{e}_{-q}$ is the complex unit vector, $\hat{e}_{\pm 1} = \mp \frac{1}{\sqrt{2}} (\hat{e}_x \pm i \hat{e}_y)$, $\hat{e}_0 = \hat{e}_z$ are the spherical basis vectors in terms of the Cartesian basis vectors $\hat{e}_x$, $\hat{e}_y$ and $\hat{e}_z$, and $\mathbf{u}_{\pm 1} = \mp \frac{1}{\sqrt{2}} (\mathbf{u}_x \pm i \mathbf{u}_y)$, $\mathbf{u}_0 = \mathbf{u}_z$ are the spherical components of $\mathbf{u}$ in terms of the Cartesian components $\mathbf{u}_x$, $\mathbf{u}_y$ and $\mathbf{u}_z$, $\alpha_v$ is the vector polarizability such that
\begin{equation}
   \alpha_v (F;\omega) = \sum_{F'} (-1)^{F+F'+1}\sqrt{\frac{6F(2F+1)}{F+1}}
   \begin{Bmatrix}
   1 & 1 & 1 \\
   F & F & F'
   \end{Bmatrix}
   \frac{\omega {|\langle J||\mathbf{d}||J' \rangle |}^2}{\hbar (\omega_{F'F}^2 - \omega^2)}(2F'+1)(2J+1)
   \begin{Bmatrix}
   J & J' & 1\\
   F' & F & I
   \end{Bmatrix}^2, \label{SupEq2}
\end{equation}
where $\omega$ is the angular frequency of light, $\omega_{F'F}$ is the resonance frequency from hyperfine level $F$ to $F'$,
$\begin{Bmatrix}
j_1 & j_2 & j_3 \\
j_4 & j_5 & j_6
\end{Bmatrix}$
is the Wigner 6-j symbol, $\hbar$ is the reduced Planck's constant, $\mathbf{d}$ is the dipole operator, $\langle J||\mathbf{d}||J' \rangle$ is the reduced matrix element, $J$ is the electron angular momentum, and $I$ is the nuclear spin momentum.

Since the frequency of our intensity-modulated light is detuned tens of GHz from the cesium D2 transition $\ket{F = 3} \rightarrow \ket{F' = 4}$, we only include the D1 and D2 transitions in our calculation, with all the relevant calculation parameters listed in Table \ref{tab1}. Therefore, the calculated fictitious magnetic field generated by a 23GHz red-detuned, $\sigma^-$ polarized and 0.87 W/cm$^2$ light is 35.6 mG.

\begin{table}[htbp]
    \centering
    \setlength{\tabcolsep}{10pt}
    \begin{tabular}{cccc}
    \toprule[1.5pt]
    \multicolumn{2}{c}{Transition} & $\omega_{F' F}/2 \pi$ (THz) & $\langle J||\mathbf{d}||J' \rangle$ (C $\cdot$ m) \\
    \midrule[1.0pt]
    \multirow{2}{1cm}{D1} & $F = 3 \rightarrow F' = 3$ & 335.12056284 & \multirow{2}{2cm}{$2.6980 \times 10^{-29}$} \\
    & $F = 3 \rightarrow F' = 4$ & 335.12173052 \\
    \\
    \multirow{3}{1cm}{D2} & $F = 3 \rightarrow F' = 2$ & 351.73054972 &  \multirow{3}{2cm}{$3.7971 \times 10^{-29}$} \\
    & $F = 3 \rightarrow F' = 3$ & 351.73070092 & \\
    & $F = 3 \rightarrow F' = 4$ & 351.73090217 & \\
    \bottomrule[1.5pt]
    \end{tabular}
    \caption{The used parameters for the calculation of $F = 3$ states' fictitious magnetic field.}
    \label{tab1}
\end{table}

Moreover, we vary the polarization as well as detuning from -200 GHz to 50GHz in experiments, observing two intriguing results. Firstly, the Feshbach resonance shift caused by linear polarized light is on the same order of magnitude as $\sigma^-$ case, which indicates that the scalar shift difference between the free-scattering state and the molecular states could not be neglected. When the light intensity is further modulated, corresponding modulation-induced resonances (including higher orders) are still observed. When the polarizations of two laser beams are linear and orthogonal to each other, no resonance is observed due to the absence of modulation.
Secondly, the shifted value decreases monotonically with the increasing detuning in blue detuned case, while it can dramatically change its sign at some detuning points in red detuned case. This behaviour may be attributed to the bound-bound transitions between the ground molecular states and the excited molecular states \cite{Ref6,Ref7}. Despite its complex details, it's sufficient for us to know that the collisional states can be distinctly shifted, so that we can realize the modulation-induced Feshbach resonance.

\textcolor{black}{\section{Heating induced by a far-detuned laser}}

\begin{figure*}[t]
    \centering
    \includegraphics[width=1\textwidth]{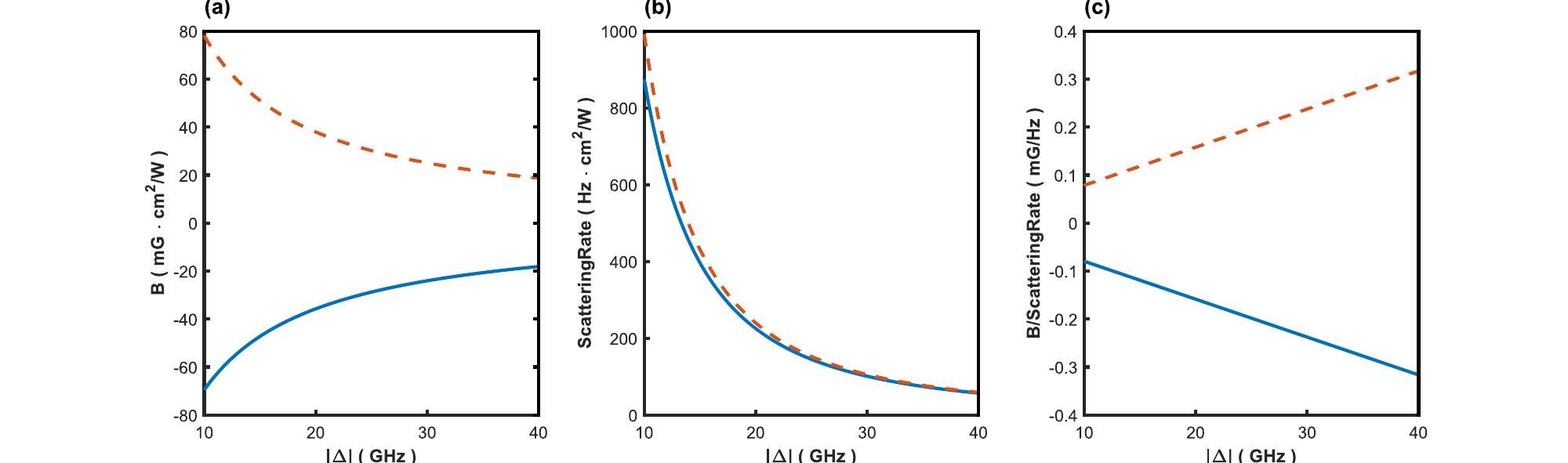}
    \caption{\textcolor{black}{The calculated fictitious magnetic field (a), scattering rate (b), ratio between fictitious magnetic field and scattering rate (c), with a $\sigma_{-}$ polarized light at the absolute detunings ranging from 10GHz to 40GHz. The blue solid curve and red dashed curve in (a)-(c) represent the blue-detuned and red-detuned cases, respectively. The two cases generate the fictitious magnetic fields with the similar absolute value but the opposite sign, causing the nearly equal scattering rate simultaneously. And in the aforementioned detuning range, larger detuning will produce larger fictitious magnetic field with the same scattering rate for both two cases.}}  
    \label{fig1}
\end{figure*}

\textcolor{black}{When a laser field is applied, the spontaneous decay of the excited states is inevitable. This process limits atoms' lifetime to milliseconds in previous studies of optically controlling the Feshbach resonances via a bound-to-bound transition \cite{Ref5,Ref6}. A substantially far-detuned laser light enables effectively shifting the magnetic Feshbach resonance and simultaneously reducing such heating to a scattering rate of $\sim$ Hz \cite{Ref1}. Here, we theoretically and experimentally investigate the heating induced by our far-detuned (tens of GHz detuning) laser light.}

\begin{figure*}[t]
    \centering
    \includegraphics[width=0.8\textwidth]{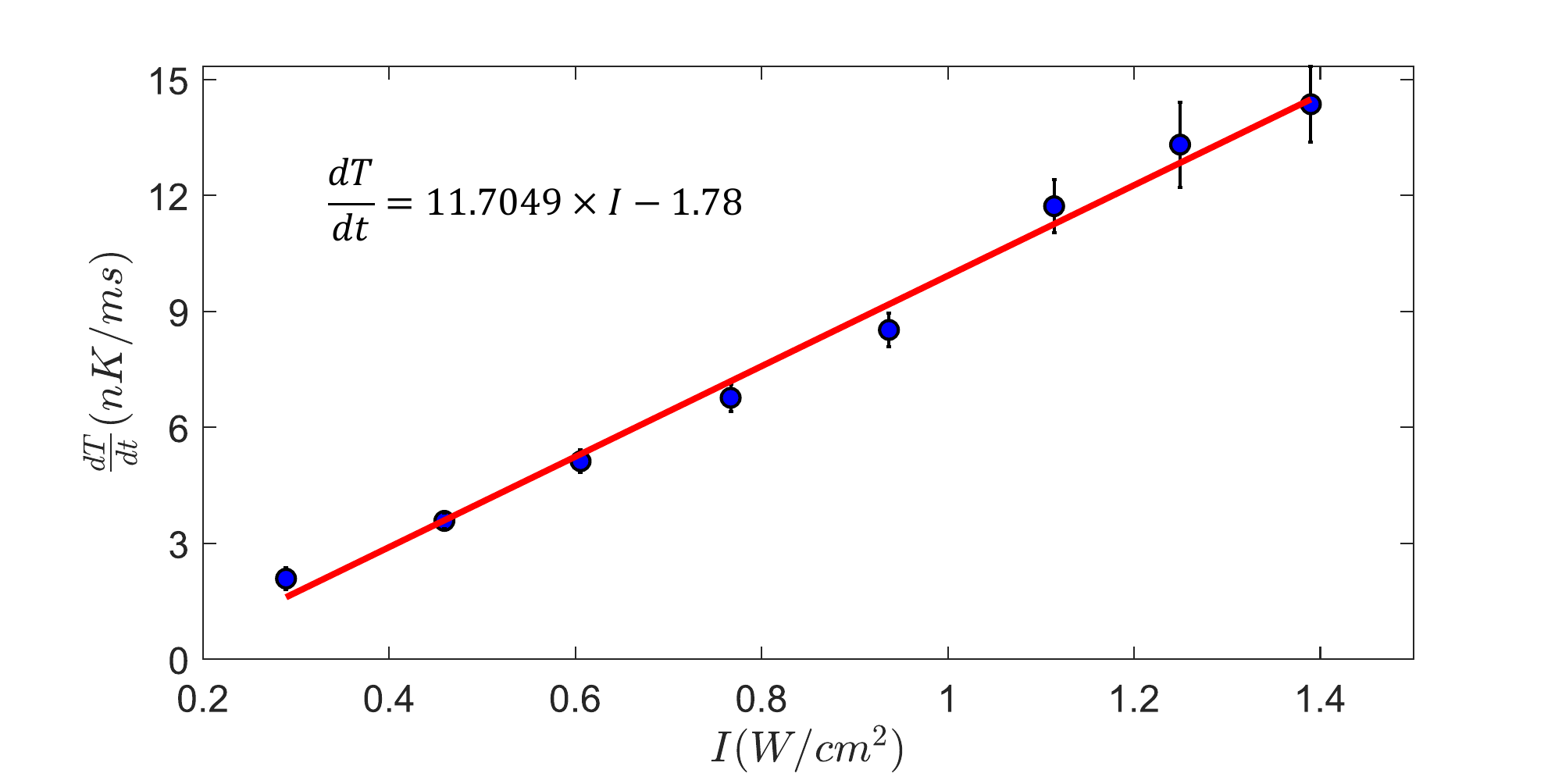}
    \caption{\textcolor{black}{The measured heating rate with a 24 GHz red-detuned $\sigma^-$ polarized laser light at different light intensity. Each data (a blue solid circle) is extracted from a linear fitting between the measured temperature and the laser light hold time. Due to the effect of evaporation cooling, the fitting (red solid line) between the measured heating rate and the light intensity, has a negative interception. Errorbar represents the standard error.}}  
    \label{heating}
\end{figure*}

\textcolor{black}{The theoretical photon scattering rate of the ground state $|F,m_F\rangle$ is \cite{Ref23}
\begin{equation}
   R_s = \frac{I \mu_0 c}{2 \hbar^2} \sum_{F' m'_{F}} \frac{|\bra{F' m'_{F}} \mathbf{u} \cdot \mathbf{d} \ket{F m_{F}}|^2}{(\omega - \omega_{F'F})^2} \Gamma_{F' m'_{F}} . \label{SupEq3}
\end{equation}
where $\Gamma_{F' m'_{F}}$ is the decay rate of the state $\ket{F' m'_F}$.
The matrix element $\bra{F' m'_{F}} \mathbf{d}_q \ket{F m_{F}}$ is decided by
\begin{equation}
    \bra{F' m'_{F}} \mathbf{d}_q \ket{F m_{F}} = (-1)^{F-1+m'_F} \sqrt{2F'+1}
    \begin{pmatrix}
    F & 1 & F' \\
    m_{F} & q & -m'_F
    \end{pmatrix}
    (-1)^{F+J'+1+I} \sqrt{(2F+1)(2J'+1)}
    \begin{Bmatrix}
    J' & J & 1 \\
    F & F' & I
    \end{Bmatrix}
    \langle J'||\mathbf{d}||J \rangle . \label{SupEq4}
\end{equation}
where $\mathbf{d}_q$ is the spherical $q$ component of the dipole operator,
$\begin{pmatrix}
j_1 & j_2 & j_3 \\
m_1 & m_2 & m_3
\end{pmatrix}$
is the Wigner 3-j symbol.}

\textcolor{black}{Similar to the last section, we only include the D1 and D2 transitions in our calculation, with $\Gamma_{D1} = 2\pi \times 4.5612$ MHz, $\Gamma_{D2} = 2\pi \times 5.2227$ MHz and the other relevant calculation parameters listed in Table \ref{tab1}. The calculated fictitious magnetic field, scattering rate, ratio between fictitious magnetic field and scattering rate, are shown in Fig. \ref{fig1}, with a $\sigma_{-}$ polarized light at different detunings.}

\textcolor{black}{To establish quantitative correspondence between the theoretical scattering rate and experimental temperature measurements, we assume that the momentum of the scattered photons is absorbed by atoms, with the resulting heating rate derived as follows
\begin{equation}
R_h=\frac{dE}{dt} \approx R_s \frac{(\hbar k)^2}{2m} \approx \frac{3k_B}{2}\frac{dT}{dt},
\end{equation}
where $T$ is is the temperature of gas, and $\hbar k$ is the momentum of a photon.} 

\textcolor{black}{We examine this estimated heating rate in experiments by preparing a cluster of $100$ nK thermal gas containing approximately $2\times 10^5$ atoms and implement temperature measurement after shinning a 24 GHz red-detuned $\sigma^-$ laser light. Subsequently, the heating rate $dT/dt$ is obtained through a linear fitting between the measured temperature and the corresponding laser light hold time. In Fig.~\ref{heating}, the measured temperature heating rate per 1 W/cm$^2$ light intensity is 11.7 nK$\cdot$ms$^{-1}/($W$\cdot$cm$^{-2}$), and it is consistent with the theoretical result of 11.0 nK$\cdot$ms$^{-1}$/(W$\cdot$cm$^{-2})$ (the corresponding scattering rate is 167~Hz/(W$\cdot$cm$^{-2}$)). This suggests that the additional heating associated with bound-to-bound transitions and photoassociation is negligible at this detuning. In experiments, significant additional heating is observed only when the detuning is very close to these transitions. Generally, the upper bound of the atom lifetime with a tens-of-GHz-detuned laser light is between the bound-to-bound transition case ($\sim$ ms) and the significantly far-detuned laser light case ($\sim$ s).}

\section{Scattering properties of modulation-induced Feshbach resonance} \label{section3}
To study the scattering properties of modulation-induced Feshbach resonances, we employ a two-channel model to simplify the problem. The system can be described by the Hamiltonian $H_0 = H_{op}\ket{op}\bra{op} + W(r)\ket{op}\bra{cl} + W(r)\ket{cl}\bra{op} + (H_{cl} +\hbar A cos\omega t)\ket{cl}\bra{cl}$, where $H_{op}$ and $H_{cl}$ denote Hamiltonian for the open channel and the closed channel, respectively, $W(r)$ is the coupling between them and A is the shaking amplitude as mentionend in the main text.

Subsequently, We take an unitary transformation $H_1(t)=U (t) H_0(t) U^\dagger (t)-i\hbar U(t) \partial U^\dagger(t)/\partial t$, and choose $U(t)$ in the diagonal form $U(t)=1 \ket{op}\bra{op} + e^{i\left[m\omega t+\frac{A}{\omega}\sin(\omega t)\right]} \ket{cl}\bra{cl}$, where m is an integer. As a result, $H_1(t)$ becomes $H_{op}\ket{op}\bra{op} + W(r) e^{-i\left[m\omega t+\frac{A}{\omega}sin(\omega t)\right]} \ket{op}\bra{cl} + W(r) e^{i\left[m\omega t+\frac{A}{\omega}sin(\omega t)\right]} \ket{cl}\bra{op} + (H_{cl} -m\hbar \omega)\ket{cl}\bra{cl}$. 

By using the expansion $e^{i\frac{A}{\omega}\sin(\omega t)}=\sum_n J_n (A/\omega)e^{in\omega t}$, where $J_n (x)$ is the $n$-th order Bessel function of the first kind, and neglecting the high-frequency terms when the incident energy $E$ approaches the resonance energy $E_{res} = \hbar \omega_{cl} - m\hbar \omega$, where $ \hbar \omega_{cl}$ is the molecular state energy in the closed channel with respect to the free-scattering states, we can simplify the Hamiltonian to $H_1(t) = H_{op}\ket{op}\bra{op} + (-1)^m W(r) J_m(A/\omega) \ket{op}\bra{cl} + (-1)^m W(r) J_m(A/\omega)\ket{cl}\bra{op} + (H_{cl} -m\hbar \omega)\ket{cl}\bra{cl}$. Under this Hamiltonian, we can determine the scattering properties by solving the  following set of coupled stationary Schr\"{o}dinger equations with the state $\varphi_{op}(r)\ket{op}+\varphi_{cl}(r)\ket{cl}$ \cite{Ref3}:

\begin{equation}
    \begin{cases}
        H_{op}\varphi_{op}(r) + (-1)^m J_m(A/\omega) W(r)\varphi_{cl}(r) =E\varphi_{op}(r)\\
         (-1)^m J_m(A/\omega)W(r)\varphi_{op}(r) + (H_{cl}-m \hbar \omega)\varphi_{cl}(r) =E\varphi_{cl}(r)\\
    \end{cases},
    \label{SHE1}
\end{equation}

where $\varphi_{op}(r)$ and $\varphi_{cl}(r)$ represent the corresponding wave function of the state of open channel and closed channel. Notably, $\varphi_{op}(r)$ has the standard asymptotic solution as:

\begin{equation}
    \varphi_{op}(r)|_{r\rightarrow \infty} \sim \frac{1}{(2\pi )^{\frac{3}{2}}}(e^{i\mathbf{k}\cdot \mathbf{r}}+f(\theta,k)\frac{e^{ikr}}{r})
    \label{scatter}
\end{equation}

Here, $f(\theta,k)$ denotes the scattering amplitude, $k$ is the incident wave vector, and $\theta$ is the incident angle between $\mathbf{k}$ and $\mathbf{r}$. And the cross section of scattering between two channels is described by $|f(\theta,k)|^2$. From the coupled Schr\"{o}dinger equations in Eq.~\ref{SHE1}, we derive the Lippmann-Schwinger equation as:

\begin{equation}
    \begin{cases}
        \ket{\varphi_{op}} = \ket{\varphi_{\mathbf{k}}^{(+)}}+(-1)^m G_{op}(E+i\eta^{0+}) J_m(A/\omega)W\ket{\varphi_{cl}}\\
        \ket{\varphi_{cl}} = (-1)^m G_{cl}(E)  J_m(A/\omega)W\ket{\varphi_{op}} \\
    \end{cases}
\end{equation}

Where $\ket{\varphi_{\mathbf{k}}^{(+)}}$ is the background scattering state, which satisfies $H_{op}\ket{\varphi_{\mathbf{k}}^{(+)}} = \frac{\hbar^2 k^2}{2\mu}\ket{\varphi_{\mathbf{k}}^{(+)}}$, where $\mu$ is the reduced mass. The operators $G_{op}$ and $G_{cl}$ represent the Green's functions of the open channel and the closed channel, respectively, and are given by:

\begin{equation}
    \begin{cases}
        G_{op}(E+i\eta^{0+})=(E+i\eta^{0+}-H_{op})^{-1}\\
        G_{cl}(E)=(E-H_{cl}+m \hbar \omega)^{-1}\\
    \end{cases}
    \label{Green_res}
\end{equation}

As shown in Eq.~\ref{Green_res}, the Green's function of the closed channel $G_{cl}$ exhibits a singularity at the resonance energy $E_{res}$, corresponding to the modulated-induced resonance discussed in the main text. Here we construct the state of resonance $\ket{\varphi_{res}}$, which satisfies the following relationship as:

\begin{equation}
    \bra{\varphi_{res}}G_{cl}(E)\ket{\varphi_{res}} = (E-E_{res})^{-1}
    \label{res}
\end{equation}

Based on this relation, we can approximate the Green's function of closed channel as:

\begin{equation}
    G_{cl}(E) \approx \ket{\varphi_{res}}(E-E_{res})^{-1}\bra{\varphi_{res}}
\end{equation}

Therefore, by eliminating $G_{cl}$, we can reformulate the coupled Schrödinger equations using the factor $D =\bra{\varphi_{res}} (-1)^m J_m(A/\omega)W\ket{\varphi_{cl}}/(E-E_{res})$ as follow:

\begin{equation}
    \begin{cases}
        \ket{\varphi_{op}} = \ket{\varphi_{\mathbf{k}}^{(+)}}+(-1)^m G_{op}(E+i\eta^{0+}) J_m(A/\omega)W\ket{\varphi_{res}}D\\
        \ket{\varphi_{cl}} = D\ket{\varphi_{res}} \\
    \end{cases}
\end{equation}

Where $(-1)^m G_{op}(E+i\eta^{0+}) J_m(A/\omega)W\ket{\varphi_{res}}D$ describes the scattering contribution to the scattering state, which can be associated with the scattering amplitude $f(\theta,k)$. Besides, in the long-range approximation, the coordinate representation of the Green's function $G_{op}$ takes the form:

\begin{equation}
    G_{op}(E,\mathbf{r},\mathbf{r'}) \sim -\frac{(2\pi)^{3/2}\mu}{2\pi \hbar^2}\frac{e^{ikr}}{r} [\varphi_{\mathbf{k}}^{(-)} (\mathbf{r'})]^{*},
    \label{G_function}
\end{equation}
where $\varphi_{\mathbf{k}}^{(-)} (\mathbf{r'}) = [\varphi_{\mathbf{-k}}^{(+)} (\mathbf{r'})]^*$.

In the low-momentum approximation, we may neglect higher partial-wave contributions to the scattering amplitude, rendering it isotropic. Furthermore, by substituting Eq.~\ref{G_function} into Eq.~\ref{scatter}, we obtain:

\begin{equation}
    f(\theta,k) = f_{bg}(\theta,k) - \frac{(2\pi)^3\mu \bra{\varphi_{\mathbf{k}}^{(-)}} (-1)^m J_m(A/\omega)W\ket{\varphi_{res}}}{2\pi \hbar^2}D \sim -a
\end{equation}

On the other hand, the background scattering amplitude $f_{bg}$ can be approximated by the background scattering length $a_{\textrm{BK}}$. Meanwhile, factor D can be represented as: 

\begin{eqnarray}
D=\frac{\bra{\phi_{res}} (-1)^m J_m(A/\omega)W\ket{\varphi_{\mathbf{k}}^{(+)}}}{E-E_{res}-\bra{\phi_{res}} J_m(A/\omega)WG_{op}(0) J_m(A/\omega)W \ket{\phi_{res}}} 
%\nonumber \\
%=\frac{\bra{\phi_{res}}W\ket{k}}{E-E_{res}+\frac{(2\pi)^3m}{4\pi\hbar^2a_{\textrm{BK}}}|\bra{\psi_{res}W\ket{k}|^2}}
.
\end{eqnarray}

Therefore, by using the definition $E_{res} = \hbar \omega_{cl} - m \hbar \omega$, the s-scattering length can be expressed in the low-energy condition as
\begin{equation}
    a = a_{\textrm{BK}}(1-\frac{\Delta_m}{-m\omega-\omega_0}),
\end{equation}
or
\begin{equation}
    \frac{1}{a} = \frac{1}{a_{\textrm{BK}}}(\frac{-m\omega - \omega_0}{-m\omega-\omega_0 - \Delta_m}),
\end{equation}
where the resonance position is
\begin{equation}
    \omega_o  = -\omega_{cl} - \bra{\phi_{res}} J_m(A/\omega)WG_{op}(0) J_m(A/\omega)W \ket{\phi_{res}} / \hbar,
\end{equation}
and the width of the resonance is
\begin{equation}
    \Delta_m = \frac{(2\pi)^3\mu}{2\pi\hbar^3a_{\textrm{BK}}}|\bra{\varphi_{res}} J_m(A/\omega)W\ket{\varphi_{\mathbf{0}}^{(+)}}|^2.
\end{equation}

\section{Formal scattering theory of modulation-induced Feshbach resonance}
Following the instance in \cite{Ref8}, we develop another formal theory to understand the modulation-induced Feshbach resonance for our two-channel model.
Starting with a Hamiltonian $H = H_0 + V(t)$ with $V(t+T) = V(t)$, where $T = \frac{2 \pi}{\omega} $, $\omega$ is the modulation frequency,  we can apply the Floquet theory \cite{Ref9} and arrive at a set of coupled equations for the Floquet mode $\ket{\phi(t)} = \sum_n e^{-i \omega_n t} \ket{\phi_n}$,
\begin{equation}
    H_0 \ket{\phi_n} + \sum_s V_{n-s} \ket{\phi_s} = (\epsilon + \hbar \omega_n) \ket{\phi_n},
    \label{A1}
\end{equation}
where $n$ is an integer, $V_n$ is Fourier expansion of $V(t)$, $\omega_n = n \omega$, $\epsilon$ is the quasi-energy of the Floquet state. The formal solution of the equation (\ref{A1}) is an infinitely coupled set of Lippmann-Schwinger equations,
\begin{equation}
    \ket{\phi_n} = \delta_{n,0} \ket{\phi_{in}} + \frac{1}{\epsilon+\hbar \omega_n - H_0} \sum_s V_{n-s} \ket{\phi_s},
    \label{A2}
\end{equation}
where $\ket{\phi_{in}}$ is the incoming state. 

Based on this result, we choose the following two-channel Hamiltonian,
\begin{equation}
    H_0 = H_{op} \otimes \ket{op}\bra{op} + H_{cl} \otimes \ket{cl}\bra{cl},
    \label{A3}
\end{equation}
\begin{equation}
    V(t) = S(t) \otimes \ket{cl}\bra{cl} + W \otimes \ket{cl}\bra{op} + W \otimes \ket{op}\bra{cl},
    \label{A4}
\end{equation}
where $H_{op}$ and $H_{cl}$ denote Hamiltonian for the open channel and the closed channel, $W(r)$ is the coupling between them and $S(t)$ is the periodically shaking term, and substitute equations (\ref{A3}), (\ref{A4}) to the equation (\ref{A2}) with the incoming state $\ket{\phi_{in}} = \ket{\varphi_{\mathbf{k}}^{(+)}}$ and the general form of $\ket{\phi_n} = \ket{\phi_n^{op}} \otimes \ket{op} + \ket{\phi_n^{cl}} \otimes \ket{cl}$, we have
\begin{equation}
    \ket{\phi_n^{op}} = \delta_{n,0} \ket{\varphi_{\mathbf{k}}^{(+)}} + \frac{1}{\epsilon+\hbar \omega_n - H_{op}} W \ket{\phi_n^{cl}},
\end{equation}
\begin{equation}
    \ket{\phi_n^{cl}} = \frac{1}{\epsilon+\hbar \omega_n - H_{cl}} \sum_s V_{n-s} \ket{\phi_s^{cl}} + \frac{1}{\epsilon+\hbar \omega_n - H_{cl}} W \ket{\phi_n^{op}}.
\end{equation}

The scattering amplitude for a transition from the incoming state to the 0th Floquet mode is determined by $\langle \bold{r} | \phi_0^{op} \rangle$ when $r \rightarrow \infty$,
\begin{equation}
    f_0 = f_{bg} -\frac{\mu( 2\pi)^3}{2\pi \hbar^2} \bra{\varphi_{\mathbf{k}}^{(-)}} W \ket{\phi_0^{cl}} \sim - a,
\end{equation}
where $f_{bg}$ is the background scattering amplitude. With $S(t) = \hbar A cos(\omega t)$ as mentioned in the main text, $V_n = \delta_{|n|,1} \hbar A $ and the resonance occurs 
when $\epsilon + \hbar \omega_m - \hbar \omega_{cl}$ approaches 0, as $\bra{\varphi_{res}}\frac{1}{\epsilon+\hbar \omega_m - H_{cl}}\ket{\varphi_{res}} = (E-\hbar \omega_{cl}+ \hbar \omega_m)^{-1}$.

\textcolor{black}{\section{Dressed atom approach of modulation-induced feshbach resonance} \label{section5}}

\textcolor{black}{In section \ref{section3}, the scattering properties are calculated based on a simplified Hamiltonian that discards the high-frequency terms near resonance. The scattering from the incoming state to $n \geq 1$ Floquet modes (inelastic channels) by absorbing $n \geq 1$ driven quanta, is actually ignored in the simplification and will make a difference especially at the modulation frequency very close to the resonance frequency. That's because once the incoming state is resonantly coupled to the bound state in the closed channel by emitting $m$ driven quanta ($m$ is a non-zero integer number, $m<0$ denotes absorption), the bound state can also be resonantly coupled to the scattering states in the inelastic channels by absorbing $n+m$ driven quanta. In analogy to the formalism applied in microwave-induced resonance \cite{Ref13} and radio-frequency-induced resonance \cite{Ref14,Ref15,Ref16}, we employ the dressed atom approach \cite{Ref10} to quantify the influence of these inelastic channels on scattering properties.}

\textcolor{black}{The quantized field description \cite{Ref10} of the semi-classical Hamiltonian $H_0 = H_{op}\ket{op}\bra{op} + W(r)\ket{op}\bra{cl} + W(r)\ket{cl}\bra{op} + (H_{cl} +\hbar A cos\omega t)\ket{cl}\bra{cl}$ is
\begin{equation}
    H = H_{atom} + \hbar \omega a^{\dagger}a + \lambda \ket{cl}\bra{cl} (a^{\dagger} + a),  \label{s27}
\end{equation}
where $H_{atom}$ is $H_{op}\ket{op}\bra{op} + W(r)\ket{op}\bra{cl} + W(r)\ket{cl}\bra{op} + H_{cl} \ket{cl}\bra{cl}$, $a$ ($a^{\dagger}$) is the annihilation (creation) operator of the modulation field, $\lambda = \frac{\hbar A}{2\sqrt{\bar{N}}}$ and $\bar{N}\gg1$. The eigenstates of its asymptotic Hamiltonian $E_{op}\ket{op}\bra{op} + E_{cl} \ket{cl}\bra{cl} + \hbar \omega a^{\dagger}a + \lambda \ket{cl}\bra{cl} (a^{\dagger} + a)$ are
\begin{equation}
    \begin{cases}
        \ket{\widetilde{cl,N}} = e^{-\frac{\lambda}{\hbar \omega}(a^{\dagger} - a)} \ket{cl}\ket{N} \\
        \ket{\widetilde{op,N}} = \ket{op}\ket{N}
    \end{cases}
    ,
\end{equation}
where $E_{op}$ is the internal energy of the open channel and is set to be zero, $E_{cl}$ is the internal energy of the closed channel and the corresponding eigenvalues of two channels are $E_{cl} + N\hbar \omega - \frac{\lambda^2}{\hbar \omega}$ ($\frac{\lambda^2}{\hbar \omega}$ is negligible), $E_{op} + N\hbar \omega$ respectively.}

\textcolor{black}{For the incoming state $\ket{\epsilon}\ket{\widetilde{op,N}}$ with collisonal energy $\epsilon$, $W(r)\ket{op}\bra{cl} + W(r)\ket{cl}\bra{op}$ couples it to the bound state $\ket{b}\ket{\widetilde{cl,N+m}}$ with the coupling strength
\begin{equation}
     \Gamma = 2 \pi |\bra{b} W(r) \ket{\epsilon}|^2 ~ |\bra{N+m} e^{-\frac{\lambda}{\hbar \omega}(a-a^{\dagger})} \ket{N}|^2 = 2 \pi |\bra{b} W(r) \ket{\epsilon}|^2  ~ |J_{m}(\frac{A}{\omega})|^2 \label{Eq26}
\end{equation}
in the limit where $N\gg 1$ \cite{Ref10}. Similarly, the bound state $\ket{b}\ket{\widetilde{cl,N+m}}$ is coupled to the scattering state $\ket{\epsilon+n\hbar \omega}\ket{\widetilde{op,N-n}}$ with the coupling strength $\gamma = 2 \pi |\bra{\epsilon+n\hbar \omega} W(r) \ket{b}|^2  ~ |J_{m+n}(\frac{A}{\omega})|^2$.  If we only retain the incoming channel $\ket{\widetilde{op,N}}$ and the closed channel $\ket{\widetilde{cl,N+m}}$, then the system is the dressed atom version of the two channel model in section \ref{section3}. Otherwise, a three channel model \cite{Ref11} can be applied to capture the main properties by including the incoming channel (partial wave) of $\ket{\epsilon} \ket{\widetilde{op,N}}$, the closed channel (partial wave) of $\ket{b}\ket{\widetilde{cl,N+m}}$ and an inelastic channel selected according to the coupling strength. When the modulation amplitude $A$ is small, the inelastic channel (partial wave) of $\ket{\epsilon+\hbar \omega} \ket{\widetilde{op,N-1}}$ ($\ket{\epsilon- m\hbar \omega} \ket{\widetilde{op,N+m}}$) is chosen for $m>0$ ($m<0$) case since its coupling strength $\gamma$ has the lowest order of $A$ among all inelastic channels, according to the asymptotic behaviour near zero of the first kind Bessel function. The inelastic channel may also be an artificial open channel whose coupling strength $\gamma$ is the summation of all inelastic channels. }

\textcolor{black}{According to \cite{Ref11}, the real and imaginary parts of the scattering length $a = \alpha - i \beta $ is then obtained as
\begin{equation}
    \alpha = a_{BK} + \frac{1}{k} \frac{\frac{1}{2}\Gamma \hbar [k + \omega_b - m\omega - \delta \omega_m]}{\hbar^2 [k + \omega_b - m\omega - \delta \omega_m]^2 + (\frac{\gamma}{2})^2 - (\frac{\Gamma}{2})^2},
\end{equation}
\begin{equation}
    \beta = \frac{1}{k} \frac{\frac{1}{4}\Gamma \gamma}{\hbar^2 [k + \omega_b - m\omega - \delta \omega_m]^2 + (\frac{\gamma + \Gamma}{2})^2},
\end{equation}
where $\omega_b$ is the binding energy of the bound state $\ket{b}$, $\Gamma$ is the coupling strength between $\ket{\epsilon}\ket{\widetilde{op,N}}$ and $\ket{b}\ket{\widetilde{cl,N+m}}$, $\gamma$ is the coupling strength between $\ket{b}\ket{\widetilde{cl,N+m}}$ and the scattering state in inelastic channel, in the limit of vanishing $k$, $\frac{1}{2}\Gamma = k a_{BK}\hbar \Delta_m $, $\gamma \sim constant$, $\delta \omega_m$ is the resonance shift $\sim A^{2|m|}$. }

\textcolor{black}{The extreme values of $\alpha - a_{BK}$ is $\pm a_{BK}\frac{\hbar\Delta_m}{\gamma}$ at $\Delta = \omega_b - m \omega - \delta \omega_m = \frac{1}{\hbar}\gamma/2$, where the imaginary parts of the scattering length is $a_{BK}\frac{\hbar\Delta_m}{\gamma}$ as well. To maintaining the modulation induced two-body inelastic loss at a weak level, large modulation frequency detuning $\Delta \gg \gamma$ is required and the following scattering length is $\alpha - a_{BK} \approx a_{BK} \frac{\Delta_m}{\Delta}$ and $\beta \approx \frac{a_{BK}}{2} \frac{\hbar\Delta_m \gamma}{(\hbar \Delta)^2}$. Thus, a small $\gamma$ is preferred to reach a significant change of the real part scattering length under above condition. For the $m= \pm 1$ cases, $\Delta_{m= \pm 1} \sim A^2$ and $\gamma_{m=1} \sim A^4$, $\gamma_{m=-1} \sim A^0$. Subsequently, when the finer frequency resolution is accessible and the detuning is $\Delta \sim A^q (2<q<3)$, the real part of scattering length for $m = 1$ case (resonance induced by emitting one driven quantum) can be significantly enhanced without
introducing catastrophic atom losses from the imaginary part, while the $m = -1$ case (resonance induced by absorbing one driven quantum) fails.}

\textcolor{black}{Multi-frequency modulation is anticipated to further suppress this unwanted modulation-induced two-body inelastic loss through destructive interference between different pathways. In a manner analogous to dissipation suppression via two-tone Floquet driving in lattice systems \cite{Ref17,Ref18} and to recent interaction control using magnetic modulation \cite{Ref19}, the undesired inelastic coupling can be tuned and suppressed by means of periodic multi-frequency modulation. In this case, Floquet theory connects the semi-classical Hamiltonian $H_0 = H_{op}\ket{op}\bra{op} + W(r)\ket{op}\bra{cl} + W(r)\ket{cl}\bra{op} + (H_{cl} +\sum_{l=1}^{L} \hbar A_l cos(l \omega t + \phi_l))\ket{cl}\bra{cl}$ to its fully quantized Hamiltonian as \cite{Ref12} $H = H_{atom} + \hbar \omega a^{\dagger}a + \sum_{l=1}^{L} \lambda_l \ket{cl}\bra{cl} (e^{-i\phi_l} (a ^{\dagger})^l + e^{i\phi_l} a^l)$, where $\lambda_l = \frac{\hbar A_l}{2 (\sqrt{\bar{N}})^l}$ and $\bar{N} \gg 1$. Solving the scattering problem for this Hamiltonian on the diagonal basis of its asymptotic counterpart will yield full information on the scattering properties. It is also possible to suppress the unfavorable inelastic coupling with non-periodic multi-frequency modulation. For instance, this may be achieved by including another bound state that is weakly coupled to the open channel scattering continuum, resembling electromagnetic-induced loss suppression in Ref. \cite{Ref20}.}

\textcolor{black}{\section{Comparison of magnetic moment spectroscopy, microwave spectroscopy, and modulation spectroscopy}}

\textcolor{black}{In this section, we present a short comparison between modulation spectroscopy using an intensity-modulated light and other binding energy measurement methods.}

\textcolor{black}{Compared to modulation spectroscopy, magnetic moment spectroscopy with multi-step transfer does not require direct intrinsic coupling between the incoming states and the measured molecular states. Therefore, magnetic moment spectroscopy allows the detection of weakly coupled higher partial-wave molecular states, for instance, the $l=8$ partial-wave molecular states in cesium \cite{Ref29}. However, modulation spectroscopy can avoid complicated transfer processes and enable the detection of those molecular states with a short lifetime. }

\textcolor{black}{In contrast to modulation spectroscopy and magnetic moment spectroscopy, microwave spectroscopy \cite{Ref29,Ref30} does not rely on any intrinsic channel coupling, but it might necessitate much higher power to provide sufficient coupling.}

\textcolor{black}{Finally, comparing laser-based modulation spectroscopy with magnetic-field-based modulation spectroscopy \cite{Ref31,Ref32,Ref33}, laser light offers more convenient high-speed modulation without the need for dedicated circuit and antenna designs. Furthermore, an intensity-modulated laser light also enables additional detection for the near-field-insensitive collisional states that can be effectively shifted by light.}